\documentclass[twocolumn,showpacs,amsmath,prb,superscriptaddress]{revtex4-2}
\usepackage{epsfig,psfrag}
\usepackage{amsbsy}
\usepackage{amsfonts}
\usepackage{bbm}
\usepackage{tabularx}
\usepackage{euscript}
\usepackage{amsfonts}
\usepackage{slashed}
\usepackage{comment}

\usepackage{amsmath,amssymb,wasysym}
\usepackage{graphicx}
\usepackage{bm}
\usepackage{braket}
\usepackage{verbatim}
\usepackage{float}
\usepackage{bbold}
\usepackage{color}
\usepackage{xcolor}
\usepackage[colorlinks=true]{hyperref}  
\hypersetup{
    unicode=false,          
    pdftoolbar=true,        
    pdfmenubar=true,        
    pdffitwindow=false,     
    pdfstartview={FitH},    
    pdftitle={Flstar},    
    pdfauthor={Bonetti, Christos, Sachdev, Pandey, Shanker, Sharma},     
    pdfsubject={},   
    pdfcreator={},   
    pdfproducer={}, 
    pdfkeywords={} {} {}, 
    pdfnewwindow=true,      
    colorlinks=true,       
    linkcolor=blue, 
    citecolor=blue,        
    filecolor=magenta,      
    urlcolor=blue           
} 

\usepackage{amsthm}
\usepackage{enumerate}
\usepackage{soul,xcolor}
\usepackage[normalem]{ulem}

\newcommand{\bfk}{\mathbf{k}}
\newcommand{\bfkt}{\mathbf{k}_{2d}}

\begin{document}
\begin{flushright}
 \href{https://arxiv.org/abs/2510.13943}{arXiv:2510.13943} 
\end{flushright}

\graphicspath{{figures/}}

\title{Yamaji effect in models of underdoped cuprates}

\author{Jing-Yu Zhao}
\affiliation{Department of Physics and Astronomy, Johns Hopkins University, Baltimore, Maryland 21218, USA}

\author{Shubhayu Chatterjee}
\affiliation{Department of Physics, Carnegie Mellon University, Pittsburgh, PA 15213, USA}

\author{Subir Sachdev}
\affiliation{Department of Physics, Harvard University, Cambridge MA 02138, USA}
\affiliation{Center for Computational Quantum Physics, Flatiron Institute, 162 5th Avenue, New York, NY 10010, USA}

\author{Ya-Hui Zhang}
\affiliation{Department of Physics and Astronomy, Johns Hopkins University, Baltimore, Maryland 21218, USA}

\date{\today}

\begin{abstract}
Recent angle-dependent magnetoresistance measurements in underdoped cuprates have revealed compelling evidence for small hole pockets in the pseudogap regime, including observation of the Yamaji effect in HgBa$_2$CuO$_{4+\delta}$
(Chan {\it et al.\/}, \href{https://doi.org/10.1038/s41567-025-03032-2}{Nature Physics {\bf 21}, 1753 (2025)}). A key distinction between theories is their predicted Fermi volumes, measured as fractions of the square lattice Brillouin zone: $p/4$ per pocket for spin density wave (SDW) versus $p/8$ for fractionalized Fermi liquid (FL*), where $p$ is the hole doping. We calculate the $c$-axis magnetoresistance $\rho_{zz}(\theta, \phi)$ within the semiclassical Boltzmann formalism for both states, and using the ancilla layer model (ALM) for FL* in a single-band Hamiltonian. The results from the $\text{FL}^*$ phase show good consistency with current experimental data. Conversely, the results for the SDW phase are highly sensitive to the ordering momentum along the $z$-direction. An ordering vector of $Q = (\pi, \pi, \pi)$ yields predictions that starkly disagree with the experiment. The only possibility for agreement within the SDW scenario is to assume an ordering momentum of $Q = (\pi, \pi, 0)$. However, even in this specific case, the SDW scenario predicts a marginally smaller Yamaji angle at $\phi=0$ than the FL* theory, and a second Yamaji peak  near in-plane angle $\phi = 45^\circ$, which was not observed in the experiment.  In reality, the Néel ordering vector is likely uncorrelated between adjacent layers, so that there is no coherent interlayer transport of hole-pocket quasiparticles in the SDW scenario, and consequently no Yamaji effect. 
Our results support the FL* interpretation of Fermi arcs in the pseudogap phase, and establish Yamaji angle measurements as a discriminatory tool between theoretical models.
\end{abstract}

\maketitle

\section{Introduction} 

The pseudogap phase plays a central role in high-$T_c$ cuprates as the parent state from which superconductivity emerges.
Despite decades of intensive research, its microscopic origin remains one of the most prominent open questions in the field~\cite{timusk_pseudogap_1999, norman_pseudogap_2005, keimer_quantum_2015}. 
A key puzzle concerns the topology of the Fermi surface, which in the pseudogap regime manifests as an open ``Fermi arc'' in photoemission and scanning tunneling microscopy, rather than a closed, large Fermi surface. 
Whether these Fermi arcs represent the surviving segments of a large Fermi surface or instead form small Fermi pockets remains highly debated. 

In principle, quantum oscillation measurements could discriminate between these scenarios by revealing the Fermi surface geometry directly. 
However, the very high upper critical fields near optimal doping make such measurements technically challenging. All observed quantum oscillations in the underdoped regime appear in the presence of charge density wave order, which complicates the interpretation of the Fermi surface.
Recent progress in high-field transport measurements, particularly $c$-axis angle-dependent magnetoresistance (ADMR) \cite{fang_admr_2022,chan_yamaji_2024}, 
provides a new opportunity to probe the Fermi pockets directly in the pseudogap phase without the presence of field-induced charge density wave order. In particular, Chan {\it et al.\/} studied the compound HgBa$_2$CuO$_{4+ \delta}$ which has single square lattice CuO$_2$ layers with as AA stacking, making it an ideal platform to study $c$-axis transport, with a simple cosine dispersion along the $c$ axis. In contrast, the earlier study on Nd-LSCO \cite{fang_admr_2022} features a staggered stacking of layers, which complicates the features and interpretation of the ADMR \cite{musser_admr_2022} (see Appendix~\ref{app:admr}).
(We also note that the bilayer structure of YBCO makes it non-ideal for ADMR.)

Chan {\it et al.\/} \cite{chan_yamaji_2024} observed the Yamaji peak in the ADMR \cite{yamaji_1989,Kartsovnikreview,Singletonreview}, which requires the cosine dispersion along the $c$ axis present only in HgBa$_2$CuO$_{4+ \delta}$. They
found that the Yamaji peak positions are consistent with small hole pockets of area $A_{\mathrm{FS}} = p/8$ (measured in units of the Brillouin zone area), rather than a large hole-like Fermi surface of area $A_{\mathrm{Large\, FS}} = (1+p)/2$, where $p$ is the hole doping level. 
In Ref.~\cite{chan_yamaji_2024}, the Fermi pocket is approximated as a simple ellipse with area $A_{\mathrm{pocket}} = p/8$, 
ignoring the Fermi-arc nature of the quasiparticle spectrum. 
As the spectral weight vanishes on the ``back side'' of the pocket, it is crucial to revisit whether the Yamaji peak can be reproduced within a more realistic microscopic model of the pseudogap state.

There have been a number of studies of the ordering patterns in underdoped HgBa$_2$CuO$_{4+ \delta}$. The intra-unit magnetic order \cite{Greven08,Greven10} has vanishing crystal momentum and so cannot, by itself, reconstruct the Fermi surface into small pockets. Charge density wave ordering has been observed at low temperatures \cite{GrevenCDWPRB}. Such ordering is expected to lead to the formation of electron pockets and an accompanying change in the sign of the Hall co-efficient, as in YBCO \cite{Taillefer07}. There is a similar change in sign of the Hall co-efficient in HgBa$_2$CuO$_{4+ \delta}$ \cite{ChanPNAS} at low temperatures. But at the higher temperatures of the Yamaji observations, the Hall co-efficient is positive \cite{ChanPNAS}, indicating the presence of hole pockets which cannot be associated with charge density wave order \cite{chan_yamaji_2024}.

In the remaining possibilities, 
one widely discussed interpretation attributes the Fermi arcs to a spin-density-wave (SDW) order with wavevector near $(\pi,\pi)$ \cite{SchmalianPines1,SchmalianPines2,abanov_sdw_2000,Tremblay04,Shen11,Chubukov23,Chubukov25}. 
In this picture, the SDW order coexists with a conventional Fermi liquid (FL) state possessing a large Fermi surface. 
The SDW order reconstructs the large Fermi surface by folding it with wavevector $(\pi,\pi)$, leading to hole pockets, each of area $p/4$.
This framework provides a satisfactory description in extremely underdoped regime where robust long-range SDW order is present~\cite{Kondo20}. 
However, its applicability at higher doping levels is less clear, particularly in the finite-temperature regime where SDW order becomes short-ranged and strongly fluctuating~\cite{Chubukov23,Greven14,Greven25}. 

An alternative scenario does not invoke static SDW order, or indeed any broken symmetry.
Instead, it assumes that small Fermi surfaces coexist with a spin-liquid-like background, forming a fractionalized Fermi liquid (FL*) phase~\cite{senthil_flp_2003,kaul_hole_2007,kaul_algebraic_2008,yang_yrz_2006,qi_pockets_2010,Mei11,Punk15,Chatterjee16,zhang_ancilla_2020,mascot_apres_2022,Pandey25,ROPP25,SSLectures25}. 
In this case, the small Fermi pocket arises from the binding of holons and spinons into gauge-neutral electron-like quasiparticles, and was predicted to have area $p/8$ \cite{kaul_hole_2007}.  A related binding has been studied in the form of `magnetic polarons' in the low density limit \cite{Grusdt18,Grusdt19}, and in the gauge theory context \cite{LeeWen96}.

Another popular approach to the pseudogap is the `phase fluctuation' theory  \cite{EmeryKivelson,Franz98,Scalapino02,Dagotto05,Berg07,Li_2010,Li11JPhys,Li11PRB,Li_2011,Sumilan17,Majumdar22,YQi23,Xiang24,yang2025}. While this can describe the photoemission Fermi arc spectra in a limited range of temperatures above the superconducting $T_c$, it does not have a simple connection to the hole pockets 
which are the focus of our attention. We also note the holon metal state \cite{PALee89,sdw09,DCSS15b,Bonetti22}, in which spinless charge $e$ quasiparticles form Fermi pockets of area $p/4$---these quasiparticles do not have coherent tunneling between layers, and so are inconsistent with ADMR observations.  

Thus we focus on the SDW and FL* phases which can be consistent with the observation of the Yamaji effect.
The quasiparticles of FL* are gauge neutral, and so can tunnel coherently between layers, as required for the Yamaji effect.
For the SDW state, coherent tunneling between layers requires that the SDW ordering correlate between layers. We will consider three dimensional SDW ordering both 
at $(\pi,\pi,\pi)$ and $(\pi,\pi,0)$, and find that only the $(\pi,\pi,0)$ case could have a Yamaji signal corresponding to the experiments. In neutron scattering experiments on HgBa$_2$CuO$_{4+ \delta}$ \cite{Greven14,Greven25}, the spin structure factor has negligible dependence on the wavevector in the interlayer direction, indicating negligible spin correlations. This is a serious issue for the SDW explanation of the Yamaji effect, but we will consider it nevertheless. 

Both the SDW and FL* approaches predict the existence of small hole-like Fermi pockets, but they differ in the resulting Fermi surface topology and pocket area. 
In the SDW scenario, the Brillouin zone is folded by $Q=(\pi,\pi)$, yielding only two inequivalent pockets per spin species, with a pocket area 
$A_{\mathrm{FL}}=p/4$. 
In contrast, in the FL* scenario, all four pockets are independent, each with area $A_{\mathrm{FL^*}}=p/8$. 
Thus, precise measurements of the Fermi pocket geometry provide a powerful means of discriminating between these two classes of theories. 

To check if ADMR can distinguish between the SDW and FL* scenarios, we calculate the $c$-axis magnetoresistance $\rho_{zz}(\theta,\phi)$ within the semiclassical Boltzmann formalism for both cases. 
A mean-field theory for the FL* state in a single band model, amenable to transport computations, is provided by the ancilla layer model (ALM) \cite{zhang_ancilla_2020}.
We show that the FL* framework  reproduces the observed Yamaji peak as a function of the azimuthal angle $\theta$, at $p=0.1$ in HgBa$_2$CuO$_{4+\delta}$ \cite{chan_yamaji_2024}.   For in-plane angle $\phi$ close to $0^\circ$, the value of the Yamaji $\theta$ is marginally smaller for the SDW theory (assuming $\mathbf Q=(\pi,\pi,0))$, but the difference can be magnified by changes in the shape of the FL* hole pocket and the precise definition of the  physical electron operator (see Appendix~\ref{app:mixing}). Moreover, there is a significant distinction between the two states at $\phi=45^\circ$. Here the SDW scenario predicts an additional peak at $\theta\approx 70^\circ$, which is absent in both the FL* calculation and in the current experiment \cite{chan_yamaji_2024}. The current experimental result at $\phi=45^\circ$ is closer to the FL* scenario than the SDW scenario (even if we assume SDW order with momentum $\mathbf Q=(\pi,\pi,0)$). A more definitive conclusion can be reached at larger $\omega_c \tau$.

Our results support the FL* interpretation of Fermi arcs in the pseudogap phase, and demonstrate that Yamaji-angle measurements provide a powerful means of distinguishing between competing theoretical scenarios of the pseudogap. We propose future experiments  to better distinguish the FL* and SDW scenario based on the data along the in-plane angle $\phi=45^\circ$ using higher magnetic fields or cleaner samples.

\begin{figure}[t]
    \centering
    \includegraphics[width=0.98\linewidth]{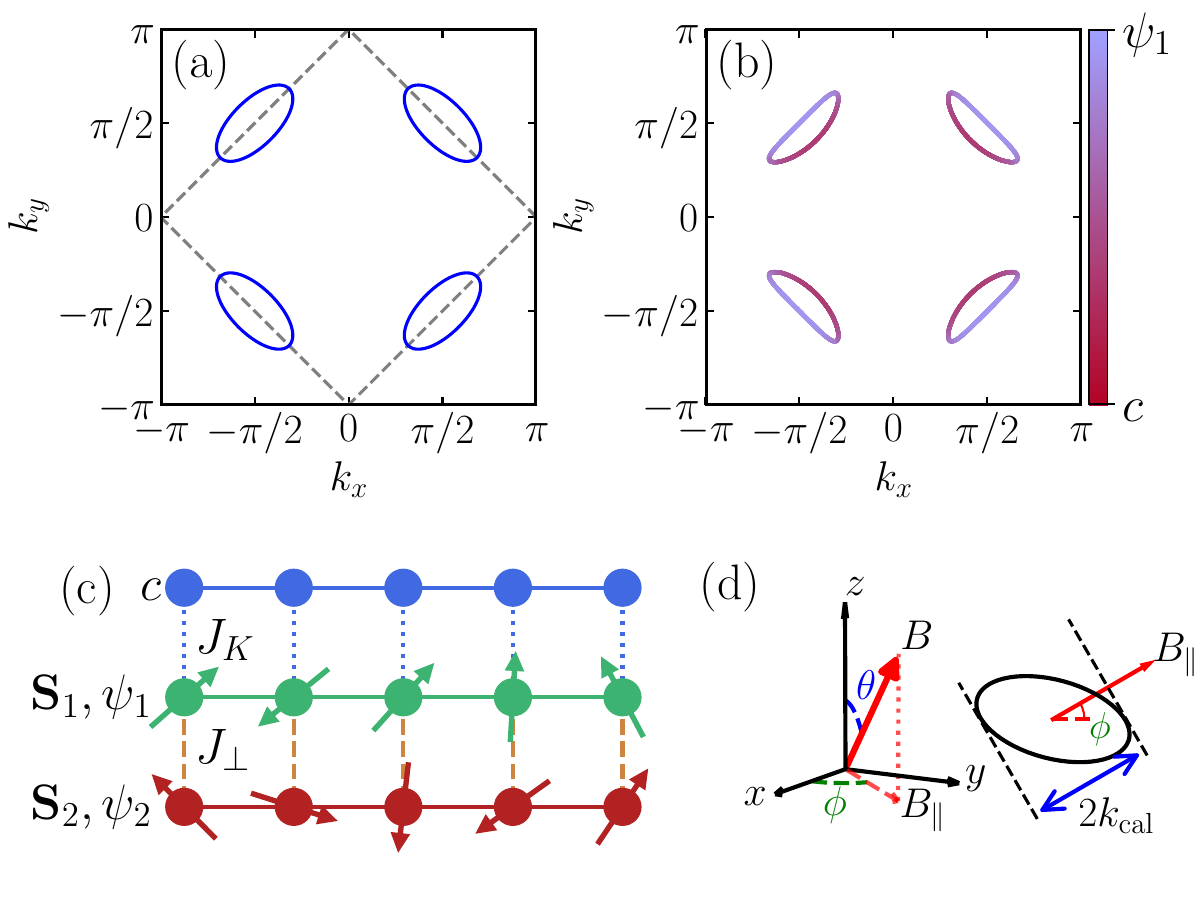}
    \caption{
    (a) and (b) shows the Fermi pocket by (a) SDW and (b) FL* by ALM at doping $p=0.1$. 
    In (b), the red color represents the contribution of the physical electron to the spectral weight on the Fermi surface, and purple color represents the contribution from the ancilla fermion $\psi_1$. 
    The dark dashed line in (a) indicate the folded Brillouin zone corresponding by SDW $(\pi,\pi)$. 
    (c) An illustration of the ancilla layer model. In the one-band model considered in this work, $c$ and $\psi_1$ are the active electrically charged degrees of freedom, while the neutral $\psi_2$ fermions decouple and do not contribute to charge transport. 
    (d) Schematic of the polar $\theta$ and azimuthal $\phi$ angles of the magnetic field $\bf B$. 
    And an illustration of the caliper momentum $k_{\mathrm{cal}}$ along the in-plange magnetic field direction $\bf B_{\parallel}$.
    }
    \label{fig:anc_fs}
\end{figure}

\section{Theoretical models for small Fermi pockets.} \label{sec:ancilla}

Here we compare different scenarios for the emergence of small Fermi pockets in the pseudogap phase.
Several competing orders—such as spin density wave (SDW), charge density wave (CDW), and pair density wave (PDW)—have been proposed in the pseudogap regime, 
where they are considered to compete with $d$-wave superconductivity \cite{Fradkin10,Hayward:2013jna,Lee14,Nie_15,Fradkin15,Pepin23,Fradkin25}. 
Specifically, we focus on two scenarios that generically lead to small hole pockets.

\textbf{SDW metal}: In this paper, we mainly consider the $\mathbf{Q}_{2d}=(\pi,\pi)$  SDW order, described by the mean field Hamiltonian:
\begin{align}\label{eqn:H_SDW}
    H_{\mathrm{SDW}}= & \sum_{\bfk,\sigma} (\epsilon_c(\bfkt)-\mu) c^\dagger_{\bfk;\sigma} c_{\bfk;\sigma} ^{} \nonumber \\
    &~~ + \sum_{\bfk,\sigma}\Delta \sigma c_{\bfk;\sigma}^\dagger c^{}_{\bfk+\mathbf{Q}_{2d} ;\sigma}+\mathrm{H.c.}\,.
\end{align}
We are using the convention that 
\begin{align}
  \bfk = (k_x, k_y, k_z)  
\end{align} in a three-dimensional momentum while 
\begin{align}
    \bfkt = (k_x, k_y)
\end{align} 
is its in-plane component,
$\epsilon_c(\bfkt)$ denotes the dispersion of free fermions, $\mu$ is the chemical potential, $\mathbf{Q}_{2d}=(\pi,\pi)$ is the SDW ordering wavevector,
and $\Delta$ is the SDW order parameter.
When $\Delta$ is finite, portions of the original Fermi surface connected by $\mathbf{Q}_{2d}$ are folded together, 
and a gap opens due to SDW ordering, leading to the formation of small closed pockets.
We assume that the SDW order is strong enough to fully gap the electron-like pockets near $(\pi,0)$.
The typical Fermi surfaces are shown in Fig.~\ref{fig:anc_fs}(a).  In the folded smaller Brillouin zone (BZ), there are two independent Fermi pockets, 
each with an area of $A = p/4$ in units of the original BZ area, where the factor of 4 in the denominator accounts for the two pockets and two spin degeneracy.  Here we have not considered the ordering in the $z$ direction, which will be important in the Yamaji effect.

\textbf{FL* phase}: The pseudogap phenomenon is robust across a range of temperatures and doping levels, 
suggesting that the gap opening and Fermi surface reconstruction may be independent of the symmetry breaking orders.
Here we consider another picture that the doped holes (with density $p$) form small pockets on top of the spin moments (with density $1$) from the parent Mott insulator. The exact nature of the spin state for the local moments does not matter much for the hole pockets as long as there is no spin-rotation symmetry breaking. 
Therefore, we assume the spin moments form a symmetric spin liquid phase. The resulting metallic state is referred fractional Fermi liquid (FL*). The description of FL* phase in a one-band model is quite challenging because it requires a two-component picture, while all electrons are identical.
We adopt the Ancilla Layer Model (ALM) \cite{zhang_ancilla_2020} for a mean field description of the FL* phase. The reader is referred to recent reviews \cite{ROPP25,SSLectures25} for further background on FL* and its ALM formulation.

In the ALM framework (as illustrated in Fig.~\ref{fig:anc_fs} (c)), we introduce two layers of ancilla fermions $\psi_1$ and $\psi_2$ in addition to the physical electron $c$. 
The physical electron $c$ hybridizes with $\psi_1$, leading to small Fermi surfaces, 
while $\psi_2$ remains decoupled and forms a spin liquid to be consistent with the Oshikawa-Luttinger theorem \cite{oshikawa_luttinger_2000}.  
In this picture, $c,\psi_1$ form the charge sector while $\psi_2$ represent the neutral spinons. 
Both $\psi_1$ and $\psi_2$ are fixed at half filling per spin, and they are projected to form a product of rung-singlet state in the end 
to encode the charge and spin degrees of freedom back to the physical band $c$. 
In this work, we focus on the charge sector and omit $\psi_2$.

The charge sector of the FL* is described by a simple mean field Hamiltonian:
\begin{equation}\label{eqn:H_anc}
\begin{aligned}
    H_{\mathrm{FL*}}=& \sum_{\bfk,\sigma}(\epsilon_c(\bfkt)-\mu)c^\dagger_{\bfk;\sigma} c_{\bfk;\sigma}^{}  
    + \Phi \sum_{\bfk;\sigma}c_{\bfk;\sigma}^\dagger \psi^{}_{1;\bfk;\sigma}+\mathrm{H.c.}\\ 
    &+\sum_{\bfk,\sigma}(\epsilon_{\psi_1}(\bfkt)-\mu_\psi)\psi^\dagger_{1;\bfk;\sigma}\psi_{1;\bfk;\sigma}^{}~,
\end{aligned}
\end{equation}
where $\epsilon_c(\bfkt)$ describes the same large Fermi surface as in Eq.~\eqref{eqn:H_SDW}, and $\epsilon_{\psi_1}(\bfkt)$ is the dispersion of the $\psi_1$ fermions. 
The chemical potentials $\mu$ and $\mu_\psi$ are introduced to fix the average density of $c$ to the physical electron density and that of $\psi_1$ to half filling, respectively.
When $\Phi$ is finite, the ancilla electrons become part of the Fermi surface and, together with $c$, form small hole-like pockets. 
In this case, all four Fermi pockets are independent, each with area $A_{\mathrm{FS}} = p/8$, where the factor of 8 arises from the four pockets and spin degeneracy.
Notably, the ancilla Fermi Hamiltonian in Eq.~\eqref{eqn:H_anc} naturally explains the Fermi arc behavior. 
As shown in Fig.~\ref{fig:anc_fs}(b), the inner side of each Fermi pocket is dominated by the physical electron $c$, 
while the outer side is contributed by the ancilla $\psi_1$.
The later is not directly visible to spectrum measurement and therefore dark in the ARPES.

\section{Yamaji effect}

While both SDW and FL* theories predict small Fermi pockets, they differ fundamentally in their microscopic interpretations and Fermi surface sizes. Unfortunately, the high transition temperatures of cuprates make direct quantum oscillation experiments challenging. We thus turn  to the Yamaji effect which operates at relatively lower magnetic fields.

For a quasi-2D material whose conducting layers lie in the $xy$ plane, 
Yamaji \cite{yamaji_1989} showed that when a strong magnetic field $\mathbf{B}$ is applied at a tilt angle $\theta$ from the $z$-axis (see Fig.~\ref{fig:anc_fs}(d)), 
the interlayer resistivity $\rho_{zz}$ oscillates as a function of the tilt angle $\theta$. 
This behavior has a purely geometric origin: the peak positions depend only on the Fermi-surface shape and are given by
\begin{equation}\label{eqn:yamaji_angle}
    c_{\mathrm{lat}} k_{\mathrm{cal}} \tan \theta_{\mathrm{Yamaji}} = \frac{3\pi}{4} + n\pi,
\end{equation}
 $c_{\mathrm{lat}}$ is interlayer direction lattice constant and $2k_{\mathrm{cal}}$ is the caliper momentum. 
It is the total span of the Fermi surface projected onto the in-plane field direction $\mathbf{B}_{\parallel}$ (see Fig.~\ref{fig:anc_fs}(d)).  For each integer $n$, there is a corresponding Yamaji angle, but the most dominant one is always at $n=0$. 
From this perspective, the Yamaji effect is an ideal method to probe Fermi-surface geometry: 
by measuring the caliper momentum along different in-plane directions azimuthal angle $\phi$, one can directly map out the Fermi-surface contour.

In Fig.~\ref{fig:anc_fs}, we illustrate the two different Fermi pockets predicted by the SDW and FL* phases respectively. 
Under the same bare hopping parameter and a finite $\Phi$, the FL* Fermi pocket is basically the inner half of the SDW pocket, 
but now the spectrum weight on the back side of the Fermi pocket is negligible because it is dominated by the ancilla $\psi_1$. 
According to \eqref{eqn:yamaji_angle}, the SDW theory always predicts a smaller Yamaji angle than FL* along all the direction, 
and the distinction becomes most significant along the $\phi=45^\circ$ direction.  
However, it is not clear whether the simple formula in  \eqref{eqn:yamaji_angle} still hold when the spectral weight $Z_{\mathbf k}$ has a strong momentum dependence. 
Therefore, we will firectly calculate the Yamaji angle using a semiclassical method based on our microscopic models in Eq.~\eqref{eqn:H_SDW} and Eq.~\eqref{eqn:H_anc}. 

\subsection{Calculation of resistivity}\label{sec:yamaji}

To compute the $c$-axis resistivity, it is necessary to include a finite interlayer hopping 
$t_z(\bfkt)$ between neighboring $\mathrm{CuO}_2$ planes. 
The corresponding Hamiltonian reads
\begin{equation}\label{eqn:tc}
    H_z = -\sum_{\mathbf k,\sigma} 2t_z(\bfkt)\,\cos(k_zc_{\mathrm{lat}}) \,c^\dagger_{0;\mathbf k;\sigma}c_{0;\mathbf k;\sigma}^{}, 
\end{equation}
where $c_{0,\mathbf{k}}$ denotes the physical electron operator that can hop between adjacent layers, and $c_{\mathrm{lat}}$ is the $z$-axis lattice constant. Notice that we now have a a full three-dimensional dispersion which depends upon $\bfk = (\bfkt, k_z)$.

In the SDW theory, $c_{0;\mathbf{k};\sigma}$ coincides with the electron operators appearing in Eq.~\eqref{eqn:H_SDW}. 
Because we care about the $z$-direction transport, the ordering momentum in $z$ direction now matters. If we assume there is a large correlation length along the $z$-direction, there are two major ansatze with momentum $\mathbf Q=(\pi,\pi,\pi)$ and $\mathbf Q=(\pi,\pi,0)$, corresponding to antiferromagnetic and ferromagnetic inter-layer spin coupling respectively. We present the results for $\mathbf{Q}=(\pi,\pi,\pi)$ in the appendix, which are very different from the current experiment. In this case the inter-layer hopping is frustrated because of the opposite spin direction between the two nearest-neighbor layers. It turns out that the $z$-direction resistivity $\rho_{zz}$ decreases with $\theta$ at small $\theta$, opposite to the trend in the experiment.  In the main text we will mainly focus on the order $\mathbf Q=(\pi,\pi,0)$ in the SDW scenario because this may be the best case to match the experiment.  However, we note that the correlation length in the $z$-direction is likely smaller than the layer spacing in real system and thus the ordering direction between two layers may be random. 
In such a scenario, the hopping of hole-pocket quasiparticles between any two layers would depend on the overlap of the in-plane SDW ordering directions of these two layers. 
The lack of SDW correlations in the $z$-direction can effectively render such hopping random, and this poses a serious challenge to the SDW explanation of the experimental observation of the Yamaji effect~\cite{Greven14,Greven25}.
Nevertheless, In the following we will focus on SDW ordering $\mathbf Q=(\pi,\pi,0)$ and leave a study of the realistic case with short range SDW order to the future work.

Within the ALM theory for the FL* phase, $c_{0;\mathbf k;\sigma}$ is generally a linear combination of two types of quasiparticles, $c_{\mathbf k,\sigma}$ and $\psi_{1;\mathbf k;\sigma}$, as the physical electron operator $c_{0;\mathbf{k};\sigma}$ is related by a canonical transformation to the operators $c_{\bfk;\sigma}$ and $\psi_{1;\bfk;\sigma}$ in the top two layers. At leading order in $1/J_\perp$, where $J_\perp$ is the rung exchange coupling between the ancilla layers (see Fig.~\ref{fig:anc_fs} (c)), this canonical transformation is local, and hence momentum independent (see Appendix A of Ref.~\cite{Random_ancilla}). So we can write
\begin{equation}\label{eqn:c0cp}
    c_{0;\bfk;\sigma} = \cos(a) \, c_{\bfk;\sigma} + \sin(a)\, \psi_{\bfk;\sigma}\,,
\end{equation}
where the mixing angle $a$ is small for large $J_\perp$; we will mainly use $a=0$ in the main text.
Furthermore, one must also consider the possible $\bfkt$ dependence of the interlayer hopping $t_z(\bfkt)$. 
For cuprates, it is commonly assumed that $t_z(\bfkt)$ takes the form
\begin{align}
    t_z(\bfkt) = t_{z,0}+t_{z,1}[\cos(k_x)-\cos(k_y)]^2/4\,, \label{tz1}
\end{align}
as discussed in Ref.~\cite{jin_interlayer_2025}. 
In what follows, we focus on the case with $t_{z,0}\neq 0$ and defer the discussion of $t_{z,1}$ to a later stage. 

For both FL* and SDW phases, we use the semiclassical method to compute the $c$-direction conductivity, 
where the quasiparticles undergo cyclotron motion in a plane perpendicular to the magnetic field. 
To this end, we use the semi-classical Chamber formula 
\begin{equation}\label{eqn:chamber}
    \sigma_{zz} = \frac{e^2 }{4\pi^3}\int_{\mathrm{FS}} d^2k \mathcal{D}(\mathbf{k})\int _{-\infty}^0 d t' v_z(\mathbf{k}(0)) v_z(\mathbf{k}(t')) e^{t'/\tau},
\end{equation}
where $\mathcal{D}(\mathbf k)$ is the density of states at the Fermi surface, $\mathbf{k}(t)$ labels different points of the path of cyclotron motion,  $\tau$ is the average scattering time 
and $v_z = \partial\epsilon(\mathbf k)/\partial k_z$ is the $z$ direction velocity of the quasi-particle on the Fermi surface. 
The interlayer resistivity is estimated as $\rho_{zz} \approx 1 / \sigma_{zz}$, since the in-plane conductivity is typically much larger that the $c$-axis conductivity.  
When the magnetic field is high enough such that the electron can go through several cycles within $\tau$, 
the velocity $v_z$ average along the cyclotron path and periodically renders zero as $\theta$ changes. 
We will use $\omega_c \tau \equiv 2\pi \tau/T_0$ to denote the strength of magnetic field or scattering time, where $T_0$ is the 
period for an electron to complete one orbit around the Fermi pocket when the magnetic field is perpendicular to the $xy$ plane.

\begin{figure}[t]
    \centering
    \includegraphics[width=0.98\linewidth]{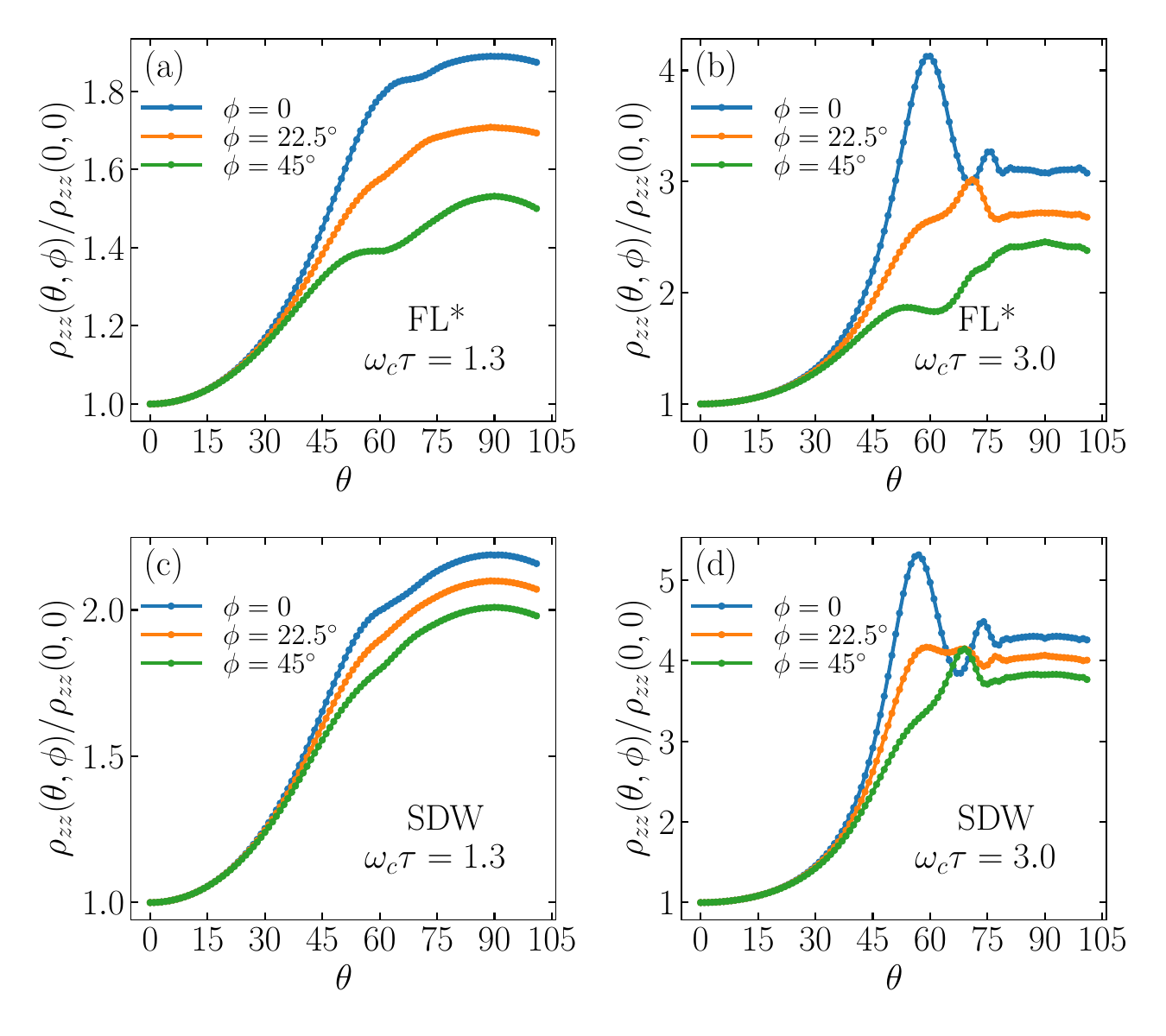}
    \caption{Calculated interlayer resistivity $\rho_{zz}(\theta, \phi)/\rho_{zz}(0, 0)$ as a function of the polar angle $\theta$ for azimuthal angles $\phi = 0^\circ$, $22.5^\circ$, and $45^\circ$.
    Panels (a) and (b) show the results obtained within FL* by ALM for $\omega_c\tau = 1.3$ and $\omega_c\tau = 3.0$, respectively, 
    with the mixing term $a=0$.
    The Yamaji angle at $\phi=0^\circ$ is extracted as $\theta\approx 59.5^\circ$. 
    Panels (c) and (d) present the corresponding results calculated within the SDW framework for $\omega_c\tau = 1.3$ and $\omega_c\tau = 3.0$. 
    The Yamaji angle at $\phi=0^\circ$ is extracted as $\theta\approx 57^\circ$. 
    }
    \label{fig:ancsdw_yamaji}
\end{figure}

\begin{figure}[t]
    \centering
    \includegraphics[width=0.98\linewidth]{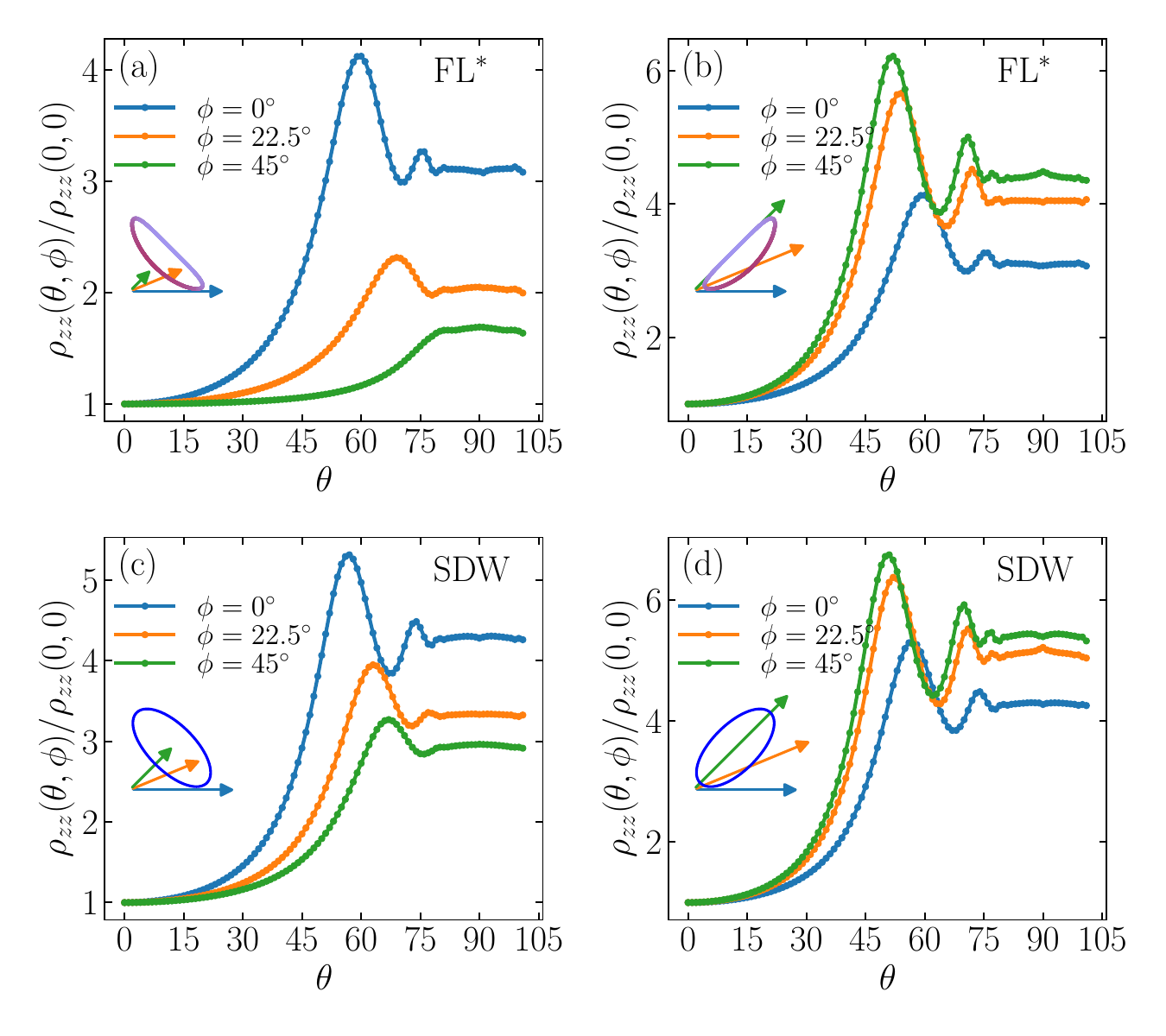}
    \caption{Comparison of the resistivity curve $\rho_{zz}$ contributed by different pockets. $\omega_c\tau=3.0$ and $a=0$ is used for all the calculations.  
    (a) and (b) show the resistivity contributed by the two pockets of the FL* theory. 
    (c) and (d) show the resistivity contributed by the two pockets of the SDW theory. 
    The corresponding pockets are shown in the insets of each figure. 
    The arrows indicate the directions of th ein plane magnetic $\mathbf B_\parallel$, 
    with the arrow lengths proportional to the corresponding caliper momentum $k_{\mathrm{cal}}$. 
    }
    \label{fig:compare}
\end{figure}

\section{Results}

We compute the resistivity as a function of the polar and azimuthal angles $(\theta,\phi)$ of the magnetic field $\mathbf{B}$ 
for both the FL* and the SDW theory, as shown in Fig.~\ref{fig:ancsdw_yamaji}. 
The calculations are performed at a hole doping level $p=0.1$ in $\mathrm{HgBa}_2\mathrm{CuO}_{4+\delta}$, 
using microscopic lattice parameters $a_{\mathrm{lat}}=3.88~\text{\AA}$ and $c_{\mathrm{lat}}=9.50~\text{\AA}$. 
For the bare hopping of electron, we take $t_c=0.22$, $t_c'=-0.034$, $t_c''=0.036$, $t_c'''=-0.007$, and $t_{z,0}=0.01$ (all in units of eV)
for both ALM and SDW theories. In the ALM theory of the FL* phase, we use the ancilla fermion $\psi$ hopping $t_\psi=0.1$, $t_\psi'=-0.03$, $t_\psi''=-0.01$, and $\Phi=0.09$ 
\cite{mascot_apres_2022}.  In the SDW, we use $\Delta=0.18$ to fully gap the electron pocket. 
The effective masses of the hole pockets are estimated to be $1.04m_e$ for the ALM theory and $3.44,m_e$ for the SDW theory, where $m_e$ is the free electron mass. 
The effective mass is extracted by comparing the cyclotron periods with that of a free hole pocket.
While these parameters are not fine-tuned to fit the experimental band structure, the calculated ADMR curve turns out to be stable against changing of the parameters.
Results are presented for both a smaller $\omega_c \tau$, 
which better mimics experimental conditions, 
and a larger $\omega_c\tau$, which provides sharper resolution of the Yamaji oscillations. 

For both theories, the resistivity curves at smaller $\tau$ show qualitative agreement with experiment: 
$\rho_{zz}$ increases with $\theta$ at small tilt angles and exhibits broad Yamaji peaks at larger $\theta$. 
Increasing $\tau$ sharpens these peaks and resolves the characteristic oscillatory structure of the Yamaji effect. 
Nevertheless, the two theories predict distinct Yamaji peak positions. 
For the FL*, the Yamaji angle is identified as $\theta \approx 59.5^\circ$ for $\phi=0^\circ$, 
while for the SDW theory it occurs at $\theta \approx 57^\circ$ for $\phi=0^\circ$. 
Experimentally, the corresponding peak is observed near $\theta \approx 63^\circ$, 
although its precise position is difficult to pin down due to the broadness of the feature. 
We also note that the Yamaji peak tends to shift slightly to larger $\theta$ for smaller $\tau$, 
likely due to the superposition of the intrinsic oscillations with the monotonic background increase in $c$-axis resistivity. 

When a finite mixing parameter $a$ is included in Eq.~\eqref{eqn:c0cp}, the distinction between the FL* and SDW descriptions becomes more pronounced.
As discussed in Appendix~\ref{app:mixing}, a finite mixing $a$ suppresses the out-of-plane quasi-particle velocity $v_z$ at the corner of the Fermi pocket, 
leading to a reduced effective caliper momentum and hence larger Yamaji angles. In the FL* scenario, a larger $\Phi$, which decides the pseudogap scale, also increases the Yamaji angle at $\phi=0^\circ$. Therefore, a larger Yamaji angle is possible in the FL* scenario with a different parameter. In contrast, in the SDW scenario, the room to tune the Yamaji angle is narrower.

Beyond the distinction in Yamaji peak positions at $\phi = 0^\circ$, a more pronounced difference appears at $\phi = 45^\circ$, 
as illustrated by the green curves in Figs.~\ref{fig:ancsdw_yamaji}(b) and (d). 
Experimentally, a broad Yamaji peak is observed near $\theta \approx 55^\circ$ for $\phi=45^\circ$, 
corresponding to the long axis of the Fermi pocket, while no clear signature of the short-axis peak has been reported. 
Our calculations reproduce this broad long-axis peak around $\theta \approx 55^\circ$ for both FL* and SDW theories. 
However, the SDW theory predicts a weker peak at $\theta\approx 55^\circ$, 
and an additional Yamaji peak near $\theta \approx 70^\circ$ at large $\tau$, 
The latter originates
from the short axis of the pocket, whereas the FL* instead exhibits a dip at the same angle. 

To clarify the origin of this discrepancy, we decompose $\rho_{zz}$ into contributions from the individual pockets, as shown in Fig.~\ref{fig:compare}. 
Along the long axis [Figs.~\ref{fig:compare}(b) and \ref{fig:compare}(d)], the resistivity curves from the FL* and SDW theories are qualitatively similar, 
with the FL* predicting a marginally larger Yamaji angle for the same reason discussed at $\phi=0^\circ$. 
In contrast, along the short axis [Figs.~\ref{fig:compare}(a) and \ref{fig:compare}(c)], the two theories produce markedly different structures: 
the SDW theory yields a Yamaji peak near $\theta \approx 70^\circ$ for $\phi=45^\circ$, 
whereas the FL* shows no corresponding peak. 
This behavior can be understood from the pocket geometry illustrated in the insets: according to Eq.~\eqref{eqn:yamaji_angle}, the Yamaji angle is determined by the Fermi momentum projected along the $\phi$ direction, $2k_{\mathrm{cal}}$.
While the projected momentum $k_{\mathrm{cal}}$ is similar for FL* and SDW along the long axis, it differs by almost a factor of two along the short axis due to the different pocket area $p/4$ in the SDW case and $p/8$ in FL* case.
Consequently, the strongest contrast between the two theories appears when $\mathbf{B}_\parallel$ is aligned with the short axis of the pocket.

For the current experimental resolution, it remains unclear whether a pronounced peak would appear at $\theta=70^\circ$ at longer quasiparticle lifetimes. 
At present, the data appears to show  a weak dip at $\theta=70^\circ$, thus is closer to the FL* scenario than the SDW scenario. 
Improved measurement precision in future high-field experiments may ultimately provide a decisive test.

\section{Discussion}

\textbf{$\psi_1$ Response to EM Field.} 
In applying the semi-classical Chambers equation [Eq.~\eqref{eqn:chamber}] to the ancilla Fermi pocket, we assume that electrons execute cyclotron orbits under the magnetic field. 
This amounts to assuming that both the $c$ and $\psi_1$ fermions respond in the same way to the external electromagnetic field.
Such an assumption is justified in two dimensions\cite{zhang_ancilla_2020}, where condensation of the bosonic field $\Phi$ locks the in-plane external magnetic field to the internal gauge field, $\mathbf{A}_\parallel = \mathbf{a}_\parallel$, via the Higgs mechanism. 
In this case, the $\psi_1$ particles are effectively coupled to the external field and therefore contribute to transport.
In a realistic three-dimensional system with finite interlayer hopping $t_z$, whether $\psi_1$ couple to the $z$ direction electromagnetic field needs re-justification. 
We note the finite mixing parameter $a$ in Eq.~\eqref{eqn:c0cp} can induce interlayer correlations of both $c$ and $\psi_1$, leading to a finite phase stiffness along the $z$ direction for the bosonic field $\Phi$. 
The stiffness along $z$ is generally much weaker than the in-plane stiffness, and this anisotropy may introduce additional broadening in the interlayer transport response.
We leave it to future to study this effect beyond the current  semi-classical calculation.

\textbf{Momentum dependence of $t_z$.} 
We have also examined the effect of $\bf k$ dependene of the interlayer hopping
$t_z(\bfkt)$ in Eq.~(\ref{tz1}),
which is symmetry-allowed and often relevant in multilayer cuprates.
As shown in Appendix~\ref{app:tz1}, when we add a $t_{z,1}$ term on top of $t_{z,0}$, the resulting Yamaji angle and overall magnetoresistance profile are nearly identical to those obtained for the isotropic case (Fig.~\ref{fig:ancsdw_yamaji}).
We therefore conclude that the anisotropic component of the interlayer hopping is unimportant in $\mathrm{HgBa}_2\mathrm{CuO}_{4+\delta}$ at low hole doping.

In general, there can also be momentum dependence in the scattering time $\tau$ at different part of the Fermi pocket.
In particular, in the FL* phase, $\tau$ for the physical electrons $c$ and for the ancilla particles $\psi_1$ may differ substantially.
Nevertheless, the Yamaji effect is expected to be robust against such anisotropy in $\tau$, provided that a closed Fermi pocket exists.  We leave it to future to analyze the effect of anisotropic scattering time.

\textbf{Short range SDW.} In the main text, we assume the spin-density-wave (SDW) has a long correlation length. Even in this idealized case, we have shown that it is in poorer agreement with the experiment than the FL* phase. In reality, the correlation length of the SDW order is known to be short in certain cuprates, like HgBa$_2$CuO$_{4 + \delta}$~\cite{Greven14,Greven25}. 
For instance, in the $z$-direction, the relative Néel ordering between adjacent layers may be uncorrelated, which would effectively randomize the inter-layer hopping. We expect the agreement with the experiment to be even poorer in this scenario, but future work is needed to rigorously model the short-range SDW case.

\textbf{ADMR in LSCO.} 
Finally, we note that angle-dependent magnetoresistance has also been measured in other cuprate materials, such as LSCO \cite{fang_admr_2022}, at a higher doping levels ($p \sim 0.21$).
In this regime, the magnetic field is not strong enough to induce the Yamaji effect.
Instead, the resistivity decreases with increasing $\theta$ at small tilt angles, and develops a peak near $\theta = 90^\circ$.
To qualitatively capture this behavior, we calculate the $c$-axis resistivity of the ancilla pocket assuming an anisotropic interlayer hopping in Eq.~(\ref{eq:t_z_term}).
For smaller $\omega_c\tau = 0.4$, $\rho_{zz}$ decreases monotonically with $\theta$ and reaches its minimum at $\theta=90^\circ$.
For larger $\omega_c\tau = 1.0$, $\rho_{zz}$ first increases to a maximum before decreasing again.
The appearance of this maximum signals the onset of a Yamaji-type oscillation,
although the first Yamaji angle is smaller than in the isotropic case due to the strong momentum dependence of $t_{z}(\bfkt)$.
We therefore predict that at sufficiently high magnetic fields, the Yamaji effect should also emerge in LSCO.

On the other hand, the resistivity peak at $\theta=90^\circ$ is absent in the present calculation because $v_z$ vanishes along the back side of the Fermi pocket,
consistent with the findings of Ref.~\cite{musser_admr_2022}.
To reproduce this feature, one needs to include a finite hybridization $a$ between the physical and ancilla layers $c$ and $\psi_1$ respectively, as discussed in more detail in Appendix.~\ref{app:admr}.

Currently, the Yamaji effect has been reported only in Hg-based cuprates, likely due to their minimal disorder. 
It is possible that the Yamaji effect could be observed in other materials under stronger magnetic fields $B$. 
Note that DFT calculations have reported conduction bands derived from Hg-O chains lying above the Fermi level. 
However, these bands, with reported energies of the order of $0.2$ eV -- 1 eV\cite{sakakibara2012,moreira2011}, are not accessible with current magnetic field strengths.

\section{Conclusion}

In summary, we have computed the $c$-axis magnetoresistance $\rho_{zz}(\theta,\phi)$ as a function of the angle $(\theta,\phi)$ of the magnetic field within the semiclassical Boltzmann framework for both the SDW and FL* scenarios of the pseudogap phase.
Using the ancilla layer model for FL*, we show that the observed Yamaji peak in HgBa$_2$CuO$_{4+\delta}$\cite{chan_yamaji_2024} is quantitatively reproduced by a small Fermi pocket of area $A_{\mathrm{FS}}=p/8$.
In contrast, the SDW scenario highly depends on the ordering in the $z$ direction. SDW with momentum $\mathbf Q=(\pi,\pi,\pi)$ qualitatively disagree with the current experiment. For SDW with momentum $\mathbf Q=(\pi,\pi,0)$ gives a reconstruction with $A_{\mathrm{FS}}=p/4$
and is less consistent with the experiment for the direction $\phi=45^\circ$ of the in-plane component of the magnetic field. For example, it predicts an additional Yamaji peak, which is absent in the current experiment. 
We conclude that FL* phase better explains the current experimental observation at $p=0.1$ and we propose future experiments to study the full doping dependence to verify the Fermi surface area $A_{\mathrm{FS}}={p}/{8}$.

\acknowledgements

We thank Mun Chan, A. Chubukov, M. Greven, N. Harrison, C.M. Jian, J. Kim, P.A. Lee, A. Nikolaenko, P.~Nosov, B. Ramshaw, T. Senthil, and L. Taillefer for valuable discussions.
SS was supported  by the U.S. National Science Foundation grant No. DMR-2245246 and by the Simons Collaboration on Ultra-Quantum Matter which is a grant from the Simons Foundation (651440, S.S.).  JYZ and YHZ were supported by the  National Science Foundation under Grant No. DMR-2237031.

\appendix

\section{Dependence of Yamaji angle on additional parameters in the FL* phase} \label{app:mixing}

\begin{figure}[t]
    \centering
    \includegraphics[width=0.98\linewidth]{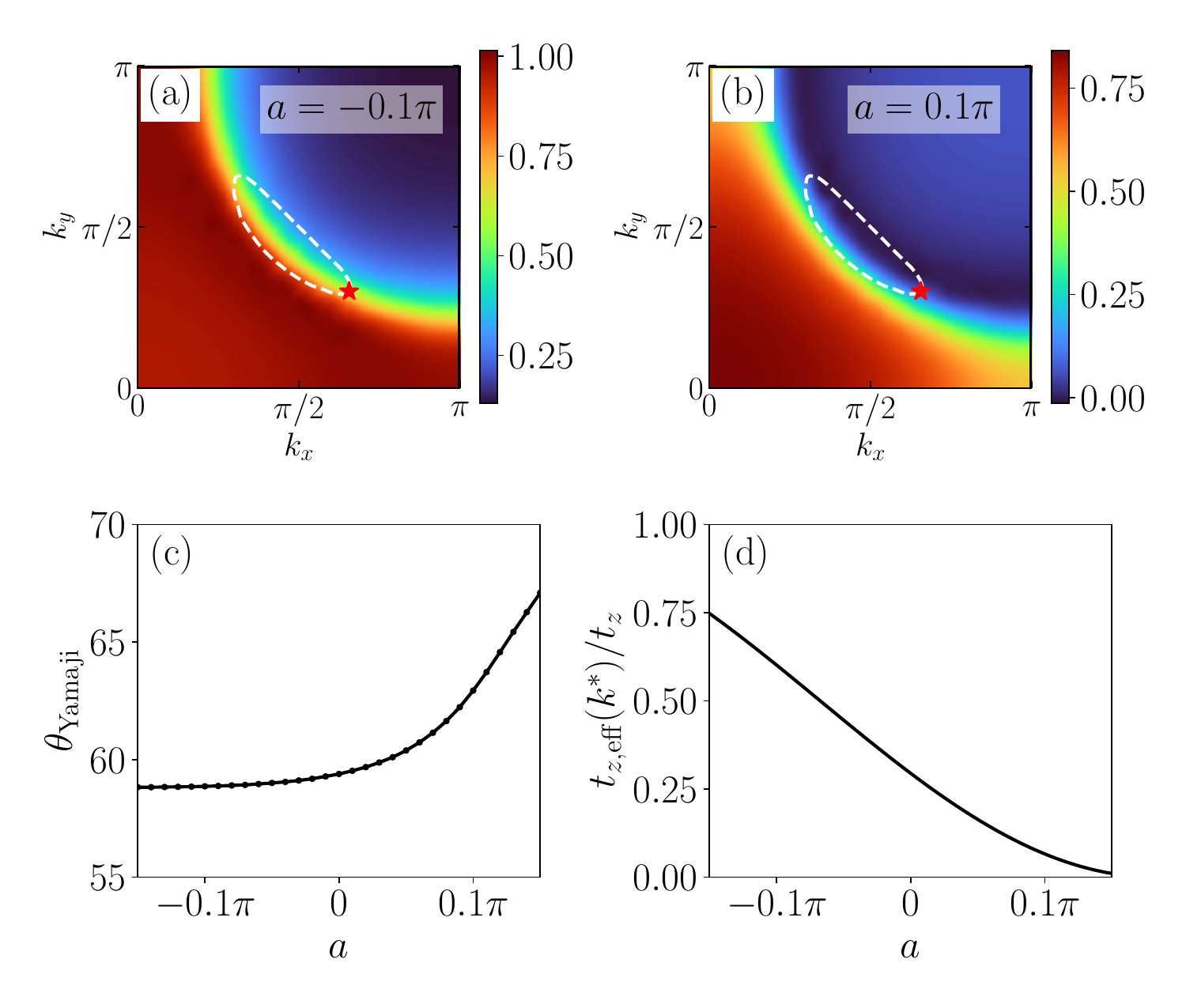}
    \caption{ Yamaji angle of ALM for different mixing angles $a$, with $a$ defined in Eq.~\eqref{eqn:c0cp} 
    (a) and (b): the effective $t_{z,\mathrm{eff}}(\bfkt)$ as a function of momentum $\mathbf k$ for $a=-0.1\pi$ and $a=0.1\pi$. 
    The dashed white line indicate the Fermi pocket and the red start marks the momentum $k^*$ of the corner of Fermi pocket. 
    (c) The Yamaji angle for $\phi=0^\circ$ as a function of $a$.
    Data points were obtained from the peak positions of the interlayer resistivity $\rho_{zz}(\theta,0)$ calculated at $\omega_c\tau=4.0$. 
    (d) The effective $t_{z,\mathrm{eff}}(\bfkt^*)$ at the corner of the Fermi pocket $\bfkt^*$}
    \label{fig:yamaji_vs_a}
\end{figure}

When a finite mixing parameter $a$ in Eq.~\eqref{eqn:c0cp} is introduced, the distinction between the SDW and FL* states becomes more pronounced.
Fig.~\ref{fig:yamaji_vs_a}(c) shows the Yamaji angle as a function of $a$, which increases monotonically with the mixing strength, especially for $a>0$. 
To gain intuition for this behavior, Fig.~\ref{fig:yamaji_vs_a} illustrates the effective interlayer hopping $t_{z,\mathrm{eff}}$ of quasiparticles for both negative and positive values of $a$
\begin{equation}
    t_{z,\mathrm{eff}} (\bfk) \equiv \frac{v_z(\mathbf{k})}{-2t_z(\bfkt)c_{\mathrm{lat}} \sin(k_zc_{\mathrm{lat}})}~, 
\end{equation}
where $v_z(\mathbf{k})=\partial\epsilon/\partial k_z$ is the out-of-plane velocity of the hybridized band formed by $c$ and $\psi_1$, 
and $-2t_z(\bfkt)c_{\mathrm{lat}} \sin(k_zc_\mathrm{lat})$ is the bare $z$ velocity computed from the interlayer hopping Eq.~\eqref{eqn:tc}.
In the limit of small $t_z$, the $k_z$-dependence of $v_z(\mathbf{k})/\sin(k_z c_{\mathrm{lat}})$ is negligible, allowing the approximation $t_{z,\mathrm{eff}}(\bfk) \approx t_{z,\mathrm{eff}}(\bfkt)$. 
For both $a<0$ and $a>0$, the out-of-plane velocity is strongly suppressed on the backside of the Fermi pocket, consistent with the fact that only one side of the pocket carries significant spectral weight $Z_\mathbf{k}$.
This behavior is qualitatively distinct from the simplified elliptical Fermi pocket used in Ref.~\cite{chan_yamaji_2024}, where a finite $v_z$ is assumed across the entire Fermi surface.

In particular, for $a>0$, the effective hopping is further reduced near the corners of the Fermi pocket, as clearly seen in Fig.~\ref{fig:yamaji_vs_a}(b) and (d).
This suppression effectively decreases the caliper momentum $k_{\mathrm{cal}}$ defined in Eq.~\eqref{eqn:yamaji_angle}, making it smaller than the geometric pocket size.
As a result, the difference between the Yamaji angles predicted by the FL* and SDW scenarios becomes more pronounced.
This observation also helps resolve a previous puzzle.
In the experimental study of Ref.~\cite{chan_yamaji_2024}, a nearly elliptical pocket with only moderate anisotropy (aspect ratio $a_{\mathrm{ellipse}}/b_{\mathrm{ellipse}} \approx 2.6$) provided good agreement with the data.
However, Fermi arcs observed in photoemission experiments often appear much flatter.
The apparent discrepancy is naturally reconciled once a finite mixing $a$ is included.

\begin{figure}[t]
    \centering
    \includegraphics[width=0.98\linewidth]{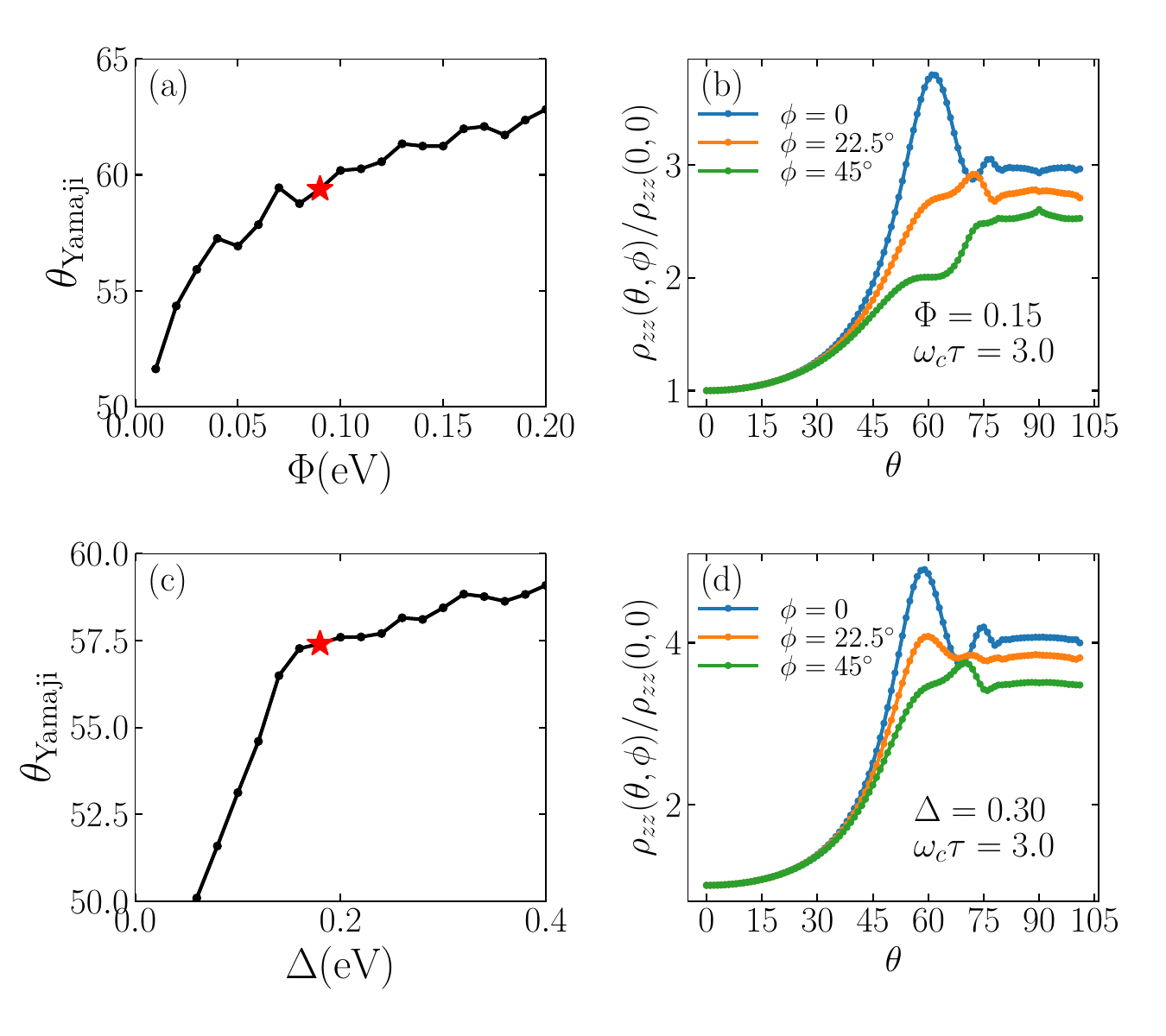}
    \caption{ Yamaji angle of FL* and SDW for different hybridization strength $\Phi$ and $\Delta$, with $\Phi$ defined in Eq.~\eqref{eqn:H_anc} and $\Delta$ defined in \eqref{eqn:H_SDW}. 
    (a) The Yamaji angle of FL* $\theta_{\mathrm{Yamaji}}$ as a function of $\Phi$ for $\phi=0^\circ$ and $a=0$, extracted from the date of $\omega_c\tau=4.0$. 
    The red star indicate the parameter we used in the main text. 
    (b) Calculated interlayer resistivity $\rho_{zz}(\theta, \phi)/\rho_{zz}(0, 0)$ for $\Phi=0.15$ as a function of the polar angle $\theta$ for azimuthal angles $\phi = 0^\circ$, $22.5^\circ$, and $45^\circ$, 
    with $\omega_c\tau = 3.0$. 
    (c) The Yamaji angle of SDW $\theta_{\mathrm{Yamaji}}$ as a function of $\Delta$ for $\phi=0^\circ$, extracted from the date of $\omega_c\tau=4.0$. 
    The red star indicate the parameter we used in the main text. 
    (d) Calculated interlayer resistivity $\rho_{zz}(\theta, \phi)/\rho_{zz}(0, 0)$ for $\Delta=0.30$ as a function of the polar angle $\theta$ for azimuthal angles $\phi = 0^\circ$, $22.5^\circ$, and $45^\circ$, with $\omega_c\tau=3.0$. 
    }
    \label{fig:yamaji_vs_Phi}
\end{figure}

There is also a strong dependence of the Yamaji angle on the hybridization strength $\Phi$.
In Fig.~\ref{fig:yamaji_vs_Phi}, we show the $\Phi$ dependence of the Yamaji angle at $\phi = 0^\circ$, which increases monotonically with $\Phi$.
A larger $\Phi$ enhances the hybridization between the $c$ and $\psi_1$ fermions, making the Fermi pocket more elliptical.
Consequently, the caliper length $k_{\mathrm{cal}}$ along $\phi=0^\circ$ becomes shorter, leading to a larger $\theta_{\mathrm{Yamaji}}$.
In the main text, we use $\Phi = 0.09~\mathrm{eV}$ as estimated in Ref.~\cite{mascot_apres_2022}, though the actual value in Hg1201 may differ slightly.
Because the Yamaji angle along a single direction is sensitive to microscopic parameters, 
only the combination of Yamaji angles from multiple directions can provide a reliable determination of the Fermi surface shape. 
This highlights the importance of our main-text discussion on the distinction between SDW and FL* at $\phi = 45^\circ$, where the difference between the two scenarios becomes most pronounced.

\section{Field dependence of $\rho_{zz}(\theta=90^\circ,\phi)$. }

\begin{figure}[t]
    \centering
    \includegraphics[width=0.98\linewidth]{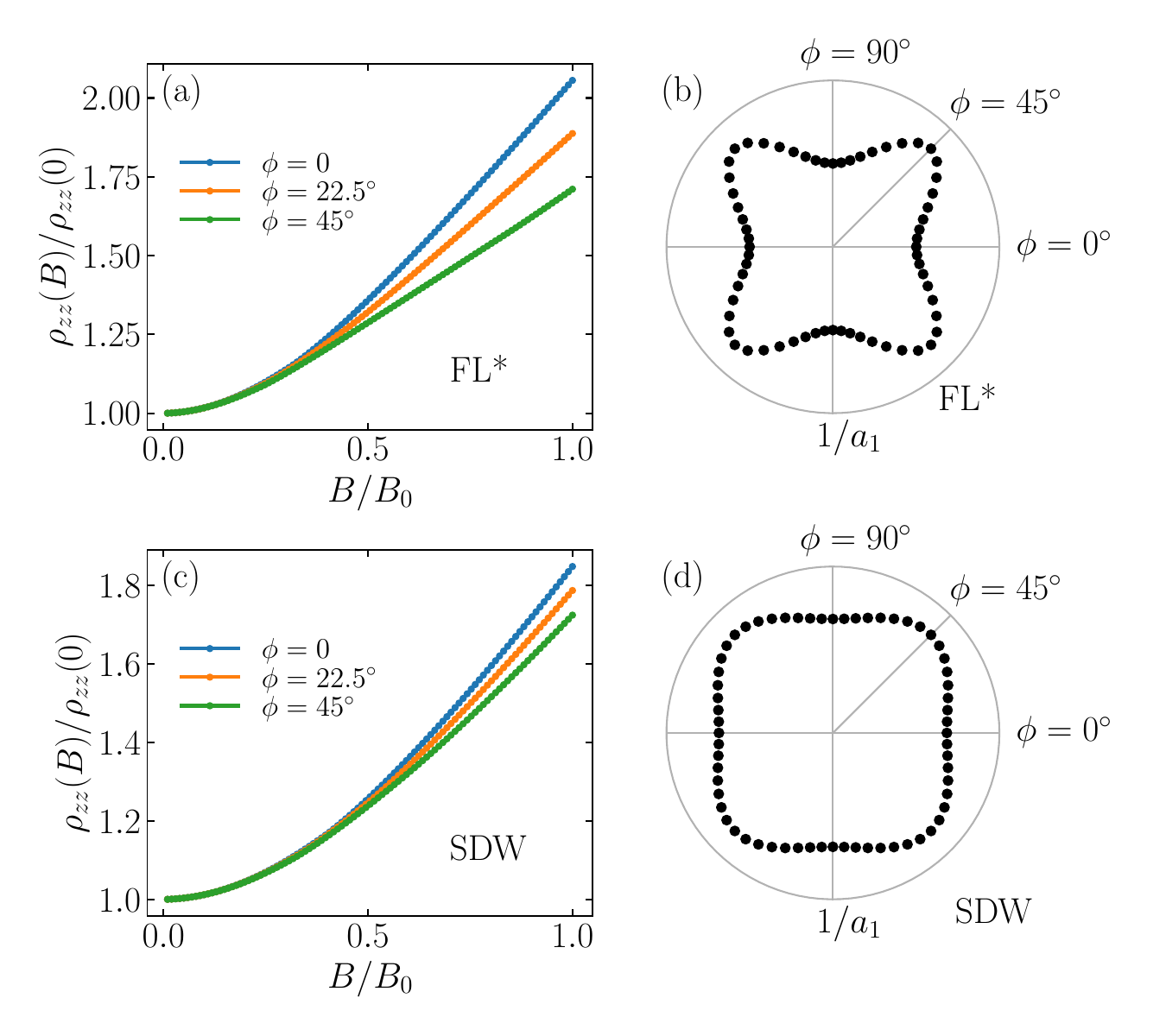}
    \caption{
    Interlayer resistivity $\rho_{zz}$ under a magnetic field for FL* (a), (b) and SDW (c), (d) states.
    (a) $\rho_{zz}(\theta=90^\circ, \phi)$ versus $B$ and $\phi$ for the FL* state, showing an approximately linear dependence $\rho_{zz} \simeq a_1 B$ at large $B$. 
    Here $B_0$ is chosen as the magnetic filed such that $\omega_c\tau=1.0$. 
    (b) Inverse slope $1/a_1$ for the FL* state as a function of $\phi$, revealing strong anisotropy with maxima near $\phi = 45^\circ$.
    (c) $\rho_{zz}(\theta=90^\circ, \phi)$ versus $B$ and $\phi$ for the SDW state, also approximately linear at large $B$.
    (d) Inverse slope $1/a_1$ for the SDW state as a function of $\phi$, showing pronounced anisotropy with maxima near $\phi = 45^\circ$. }
    \label{fig:ancsdw_sigma90}
\end{figure}

The field dependence of $\rho_{zz}$ at $\theta = 90^\circ$ 
also reflects the pocket geometry. 
At high fields, we find $\rho_{zz} \propto B$ with $\rho_{zz} = a_1 B$, 
and the inverse slope $1/a_1$ shows a pronounced dependence on $\phi$: 
for both theories, $1/a_1$ reaches a maximum at $\phi = 45^\circ$ 
and a minimum at $\phi = 0^\circ$, in qualitative agreement with experiment.

\section{Form factor of inter-layer hopping}\label{app:tz1}

\begin{figure}[t]
    \centering
    \includegraphics[width=0.95\linewidth]{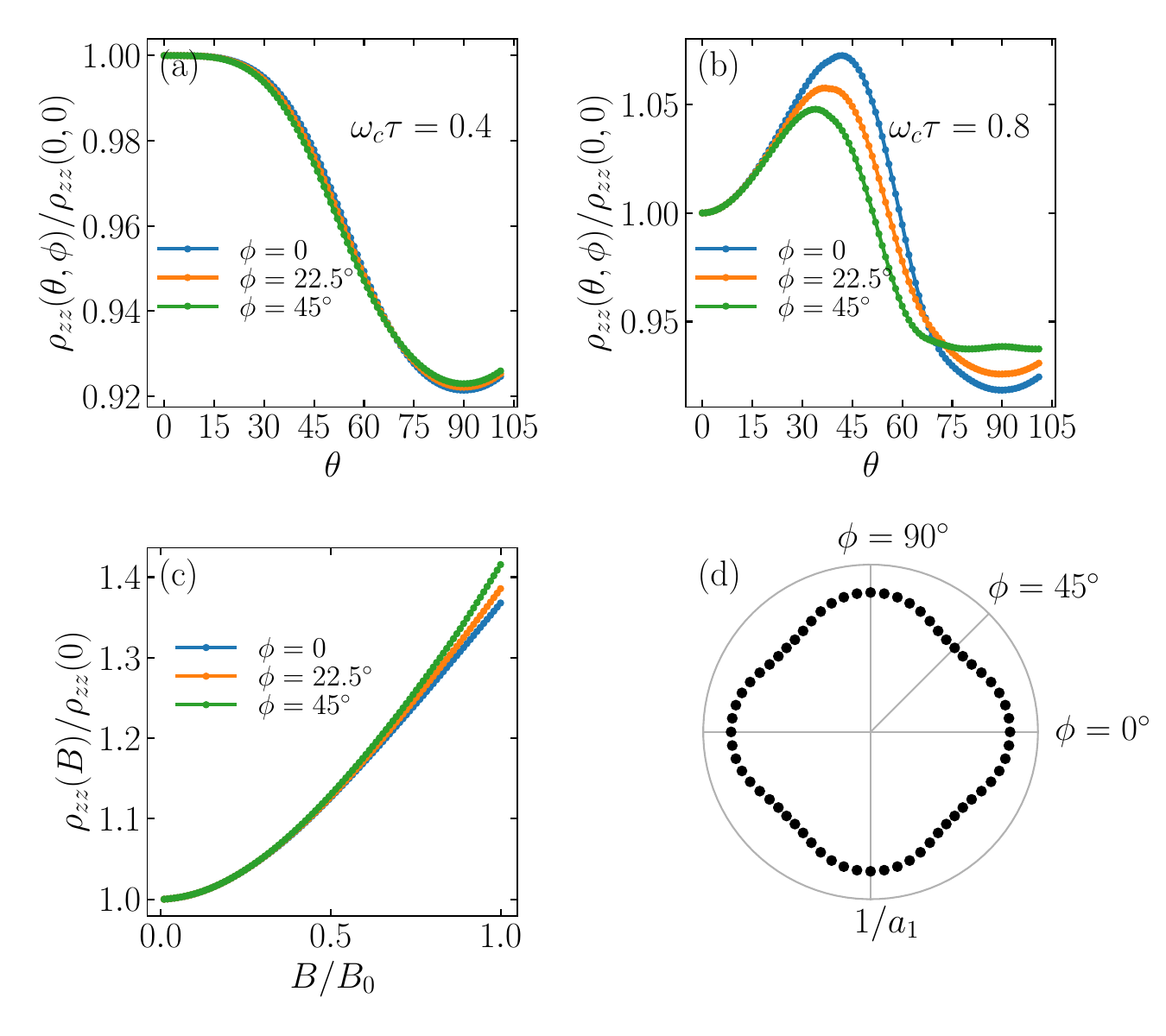}
    \caption{Resistivity $\rho_{zz}$ calculated by the ancilla pocket with $t_z(\mathbf k) = t_{z}[\cos(k_x)-\cos(k_y)]^2/4$, as a function of $\theta$ and $\phi$ for (a) $\omega_c\tau=0.4$ and (b) $\omega_c\tau=0.8$. 
    (c) $\rho_{zz}(\theta=90^\circ,\phi)$ as a function of $B$ and $\phi$. 
    Here $B_0$ is chosen as the magnetic field such that $\omega_c\tau=1.0$. 
    (d) The inverse slope $1/a_1$ as a function of $\phi$. }
    \label{fig:anct1_yamaji}
\end{figure}

\begin{figure}[t]
    \centering
    \includegraphics[width=0.95\linewidth]{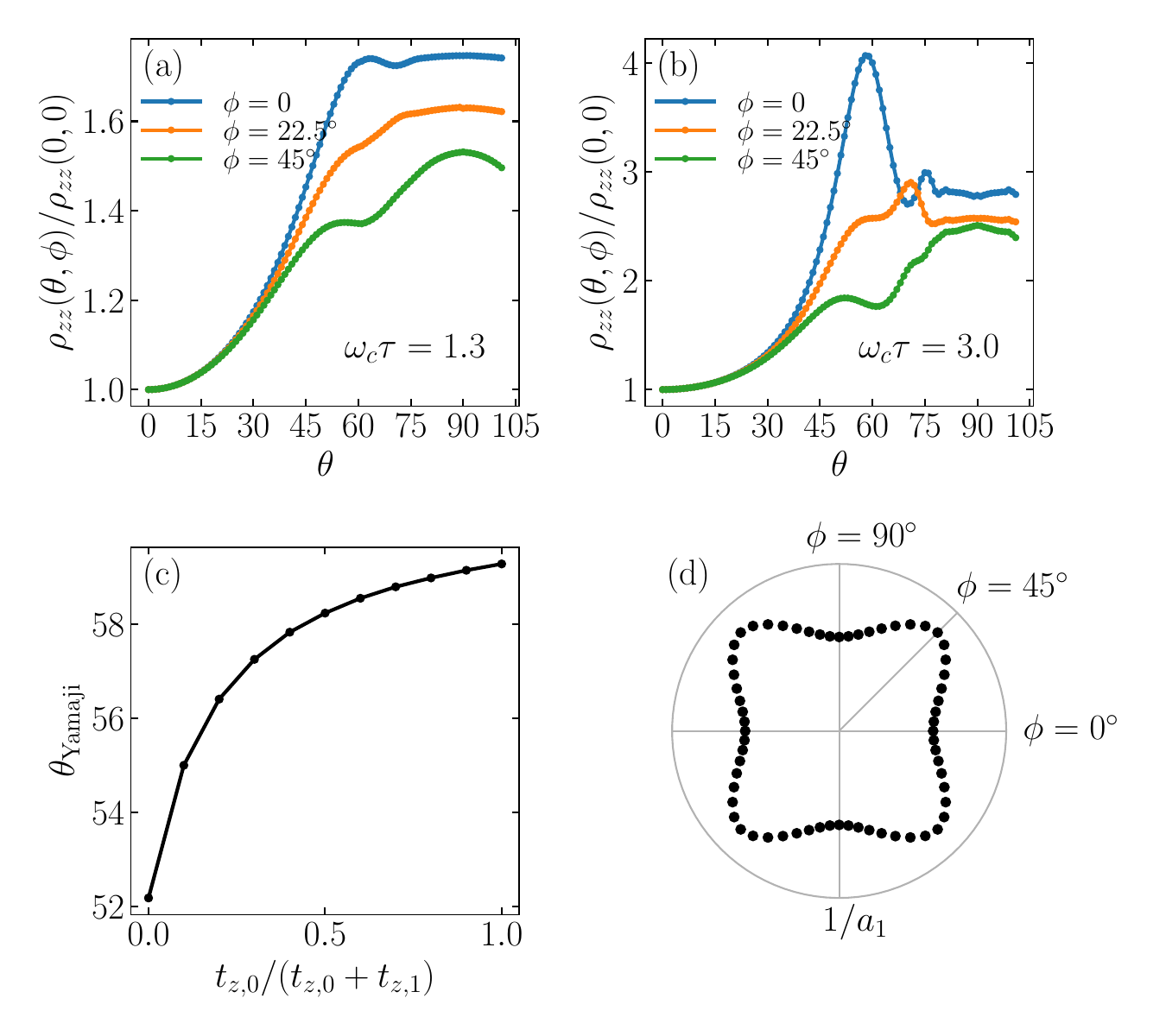}
    \caption{Resistivity $\rho_{zz}$ calculated by the ancilla theory with $t_z(\mathbf k) = t_{z}(1 +[\cos(k_x)-\cos(k_y)]^2/4)$, as a function of $\theta$ and $\phi$ for (a) $\omega_c\tau=1.3$ and (b) $\omega_c\tau=3.0$.  
    (c) Extracted Yamaji angle $\theta_{\mathrm{Yamaji}}$ as a function of the uniform component percentage. 
    (d) The slope $1/a_1$ as a function of $\phi$. }
    \label{fig:anct01_yamaji}
\end{figure}

\begin{figure}[t]
    \centering
    \includegraphics[width=0.95\linewidth]{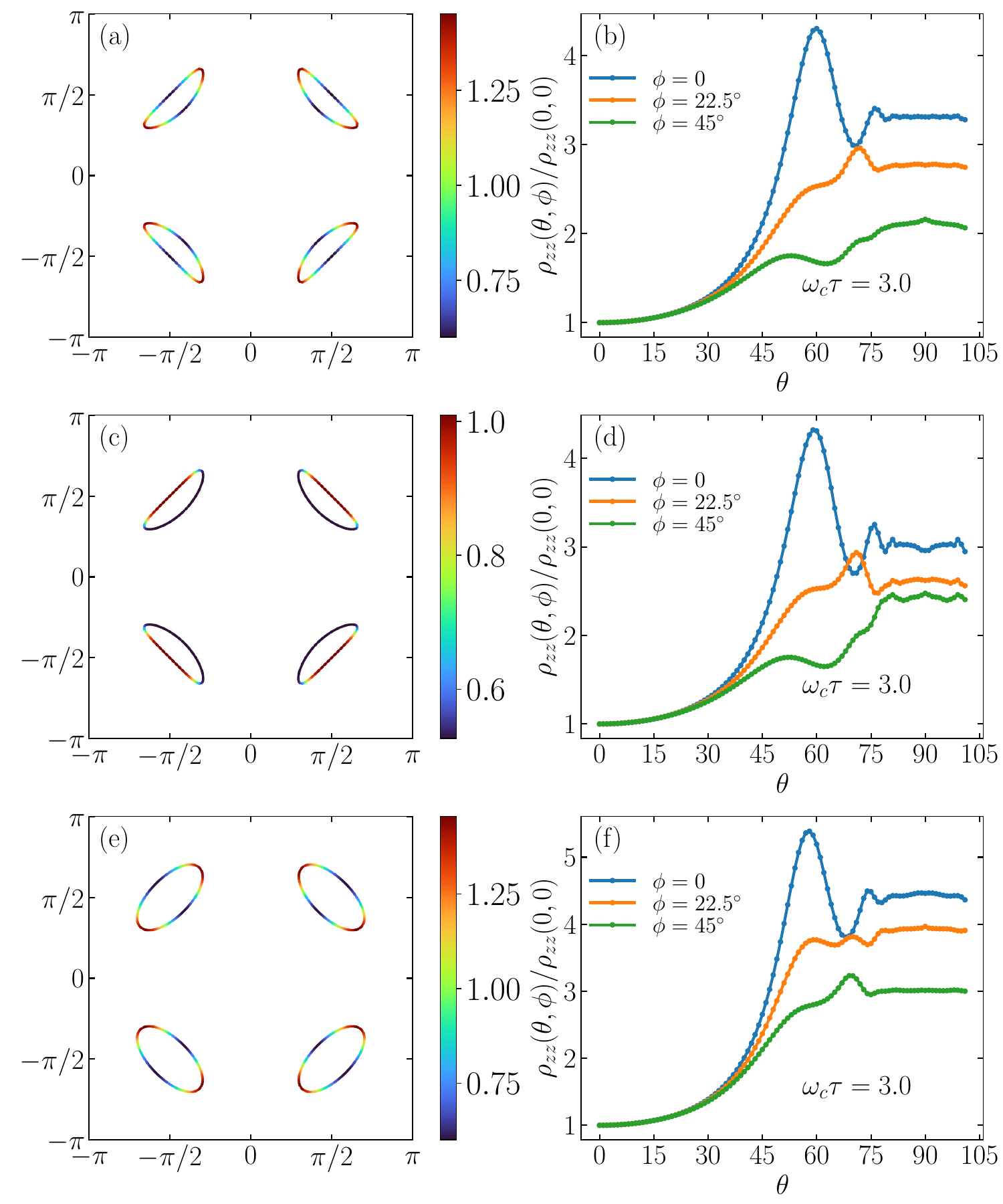}
    \caption{Out-of-plane resistivity $\rho_{zz}$ within the ALM framework [(a)-(d)] and SDW framework [(e)-(f)] for various scattering rate anisotropy models. 
    (a) Momentum-space distribution of the scattering rate $\tau_0/\tau(\mathbf{k}) = 0.15 + 0.5\cos^2\phi_\mathbf{k}$ for the FL* pocket, where $\phi_{\mathbf{k}}$ denotes the azimuthal angle. 
    The color scale represents the magnitude of $\tau_0/\tau(\mathbf{k})$.(b) Angular dependence of the corresponding resistivity $\rho_{zz}$ derived from the scattering profile in (a).
    (c) Distribution of the scattering form factor $\tau_0/\tau(\mathbf{k}) = 1/[1- 0.9\tanh (k_d/0.01)]$ for the FL* pocket, where $k_d$ is the momentum distance to the folded Brillouin zone boundary.
    (d) Calculated resistivity $\rho_{zz}$ as a function of $\theta$ for the scattering model defined in (c).
    (e) Momentum-space distribution of the scattering rate $\tau_0/\tau(\mathbf{k}) = 0.15 + 0.5\cos^2\phi_\mathbf{k}$ for the SDW pocket. 
    (f) Angular dependence of the resulting resistivity $\rho_{zz}$ for the SDW pocket case.
    }
    \label{fig:ancfrom}
\end{figure}

\begin{figure}[t]
    \centering
    \includegraphics[width=0.98\linewidth]{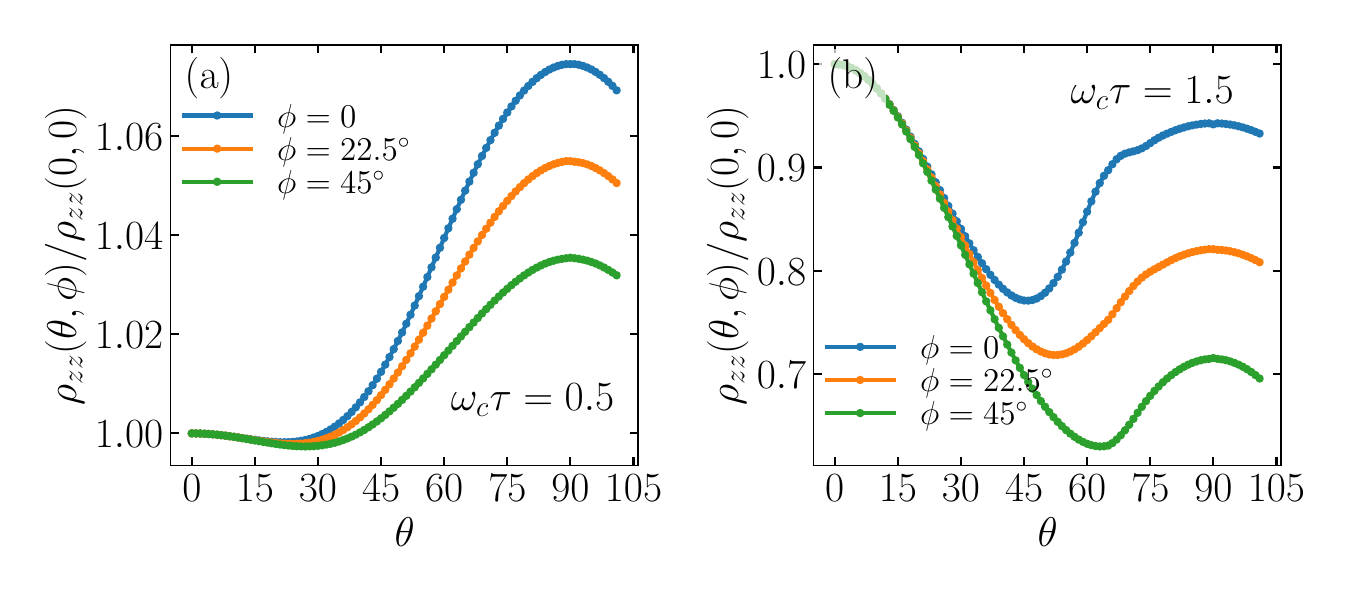}
    \caption{
        Resistivity $\rho_{zz}$ calculated within SDW theory with $\mathbf{Q}=(\pi,\pi,\pi)$,
        shown as a function of $\theta$ and $\phi$ for (a) $\omega_c \tau = 0.5$ and (b) $\omega_c\tau=1.5$.
    }
    \label{fig:sdw_pipipi}
\end{figure}

\begin{figure}[t]
    \centering
    \includegraphics[width=0.98\linewidth]{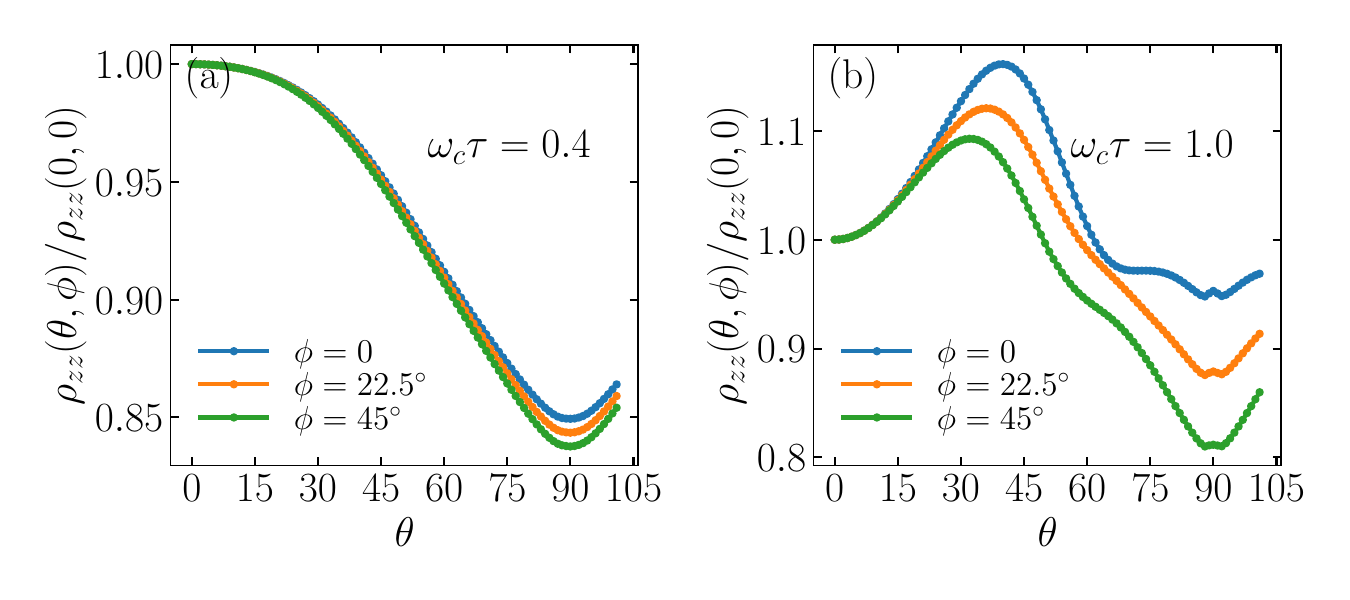}
    \caption{
        Resistivity $\rho_{zz}$ calculated within FL* for LSCO parameters at doping $p = 0.21$, with interlayer hopping Eq.~\eqref{eq:t_z_term}, 
        shown as a function of $\theta$ and $\phi$ for (a) $\omega_c \tau = 0.4$ and (b) $\omega_c\tau=1.0$.
    }
    \label{fig:anc_admr}
\end{figure}

\begin{figure}[t]
    \centering
    \includegraphics[width=0.98\linewidth]{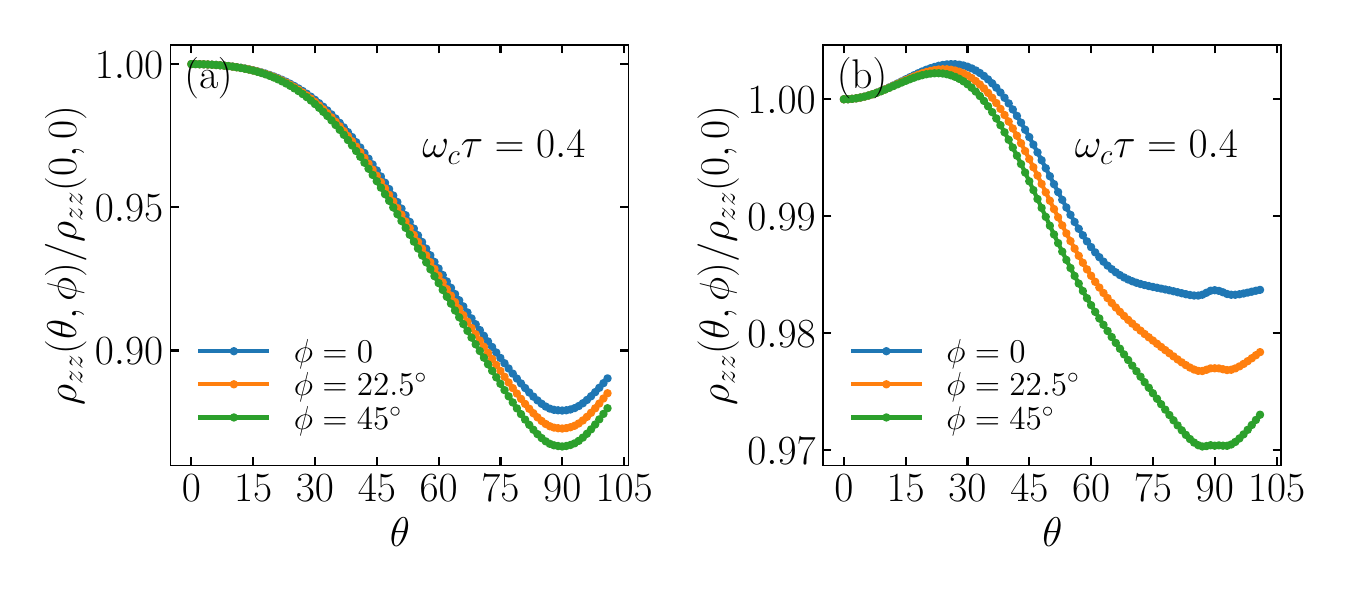}
    \caption{
        Resistivity $\rho_{zz}$ calculated within FL* for LSCO parameters at doping $p = 0.21$, with interlayer hopping Eq.~\eqref{eq:t_z_term}, 
        shown as a function of $\theta$ and $\phi$ for $\omega_c \tau = 0.4$.
        Panels (a) and (b) correspond to
        (a) $c_0 = \cos(\pi/10) c_{\mathbf{k}} + \sin(\pi/10) \psi_{\mathbf{k}}$, and
        (b) $c_0 = \cos(\pi/5) c_{\mathbf{k}} + \sin(\pi/5) \psi_{\mathbf{k}}$, respectively.
    }
    \label{fig:anc_admr_psitz}
\end{figure}

In this appendix, we examine the effect of two different forms of interlayer hopping, parameterized as
\begin{equation} \label{eq:t_z_term}
    t_{z} (\bfkt) = t_{z,0} + t_{z,1}[\cos k_x - \cos k_y\bigr]^2/4.
\end{equation}
In the main text, for simplicity, only the uniform component $t_{z,0}$ was included.
However, both $t_{z,0}$ and $t_{z,1}$ are symmetry-allowed and may coexist in realistic materials.

Figure~\ref{fig:anct1_yamaji} shows the interlayer resistivity calculated using only the anisotropic term $t_{z,1}$.  
The resulting $\rho_{zz}$ displays a qualitatively different behavior from that in the main text Fig.~\ref{fig:ancsdw_yamaji}.
For small $\tau$, the resistivity decreases monotonically with increasing $\theta$ and reaches a minimum at $\theta = 90^\circ$,
similar to the behavior reported for LSCO.  
For larger $\tau$, Yamaji oscillations appear as $\theta$ increases, but the corresponding Yamaji angles are significantly smaller than in the isotropic case.
We also examined the field dependence of $\rho_{zz}(B)$ at $\theta = 90^\circ$, and found that neither its overall trend nor its angular dependence matches experimental observations.

In contrast, when both terms are included with equal weight,
\begin{equation}
    t_{z} (\bfkt)= t_{z,0}( 1 + [\cos k_x - \cos k_y\bigr]^2/4),
\end{equation}
with $t_{z,0}=0.005$eV. The calculated resistivity curves [Fig.~\ref{fig:anct01_yamaji}] are nearly identical to those obtained using only the uniform hopping $t_{z,0}$ (Figs.~\ref{fig:ancsdw_yamaji} (a) and (b) in the main text).
We therefore conclude that in $\mathrm{HgBa}_2\mathrm{CuO}_{4+\delta}$, the uniform interlayer hopping $t_{z,0}$ provides the dominant contribution, provided its magnitude is comparable to that of the anisotropic term $t_{z,1}$.

\section{Form factor of scattering time $\tau$. }

In the primary calculations presented in the main text, we assumed a uniform scattering time $\tau$ across the entire Fermi pocket. 
However, because quasiparticles derive their spectral weight from different components at various segments of the Fermi surface, 
the scattering rate is expected to exhibit momentum dependence. 
Specifically, near the corners of the Fermi pockets, quasiparticles undergo strong scattering mediated by the bosonic field $\Phi$ within the ALM framework, or by the order parameter $\Delta$ in the SDW theory.

In this appendix, we examine the robustness of our results by considering two distinct momentum-space distributions for the scattering time $\tau(\mathbf{k})$, as illustrated in Fig.~\ref{fig:ancfrom}.
First, we consider an anisotropic scattering rate that reaches its maximum at the pocket corners, defined by:
\begin{equation}
    \tau_0/\tau(\mathbf{k}) = 0.15 + 0.5\cos^2\phi_\mathbf{k}~, 
\end{equation}
where $\phi_{\mathbf{k}}$ denotes the azimuthal angle of $\mathbf{k}$.
Second, specifically for the FL* phase, we implement a distribution where the scattering rate is higher for the ancilla fermions compared to the conventional $c$-fermions:
\begin{equation}
    1/\tau(\mathbf{k}) = 1/(1- 0.9\tanh (k_d/0.01))~,
\end{equation}
where $k_d$ represents the momentum-space distance from $\mathbf{k}$ to the folded Brillouin zone boundary. 
We find that in both models, and for both the FL* and SDW frameworks, the specific distribution of the scattering time $\tau(\mathbf{k})$ has a negligible impact on the overall results. 
Key features, including the Yamaji angles and the qualitative distinctions between the FL* and SDW phases, remain robust against scattering anisotropy.

\section{SDW with $\mathbf{Q}=(\pi,\pi,\pi)$. }

In considering the $c$-axis resistivity within the SDW framework, it is important to carefully account for the $c$ component of the SDW folding momentum. 
While the spin correlation along $c$ is generally antiferromagnetic, it is much weaker than the in-plane correlations. 
In the main text, we focus on the case with $\mathbf{Q} = (\pi, \pi, 0)$.
Here, we also examine the alternative possibility that $\mathbf{Q}=(\pi,\pi,\pi)$. 

The resulting ADMR curve is shown in Fig.~\ref{fig:sdw_pipipi}, which differs qualitatively from the experimental observation. 
The ADMR curve decreases as a function of $\theta$ for small $\theta$. 
This behavior arises from the folding of the interlayer hopping $t_z$, which causes the inner and outer halves of the SDW pocket acquire opposite signs of $t_z$,  
effectively introduceing an anisotropy in the interlayer hopping. 
At higher magnetic fields, a Yamaji peak emerges at a larger angle, but the overall lineshape remains inconsistent with experiment.

\section{ADMR in LSCO}\label{app:admr}

In this appendix, we show the magnetoresistance curve calculated by ALM in the LSCO sample\cite{fang_admr_2022} at doping $p=0.21$,  
with $a_{\mathrm{lat}}=3.75~\text{\AA}$ and $c_{\mathrm{lat}}=6.60~\text{\AA}$. 
Here we follow Ref.~\cite{fang_admr_2022} to choose $t = 0.16 $ eV, 
$t'/t = -0.1364$ and $t''/t =  0.0682$ for the $c$ electron, and choose
$t_\psi/t = -1$, $t_\psi'/t = 0.1364$ and $t_\psi''/t =  0.0682$ for the $\psi$ electron, 
and a hybridization $\Phi/t = 0.006$. 
The interlayer hopping is taken as 
\begin{equation} \label{eq:t_z_term}
    t_{z} (\bfkt) =\frac{1}{4} t_{z,1}[\cos k_x - \cos k_y\bigr]^2\cos\frac{k_x}{2}\cos\frac{k_y}{2}~,
\end{equation}
where $t_{z,1}/t=0.0325$. To account for the interlayer shift of neighboring $\text{CuO}_2$ planes in LSCO, the $t_z$ term is further multiplied by a factor of $\cos(k_x/2)\cos(k_y/2)$.
The calculated resistivity is shown in Fig.~\ref{fig:anc_admr}. 
For smaller $\omega_c\tau = 0.4$, $\rho_{zz}$ decreases monotonically with $\theta$ and reaches its minimum at $\theta=90^\circ$.
For larger $\omega_c\tau = 1.0$, $\rho_{zz}$ first increases to a maximum before decreasing again.
The appearance of this maximum signals the onset of a Yamaji-type oscillation,
although the first Yamaji angle is smaller than in the isotropic case (with $t_{z,0}$ dominated) due to the strong momentum dependence of $t_{z}(\bfkt)$.

Experiment \cite{fang_admr_2022} also revealed a pronounced peak in $\rho_{zz}$ near $\theta = 90^\circ$.
Such a feature is absent in the FL* pocket, where the backside carries a vanishing quasiparticle weight $Z_k$.
Here we show that this experimental feature can be partially captured by introducing a finite $z$-direction hopping of the ancilla fermion $\psi$,
arising from the finite mixing between $c$ and $\psi$ in Eq.~\eqref{eqn:c0cp}.
The calculated $\rho_{zz}(\theta)$ for $a = \pi/10$ and $a = \pi/5$ are shown in Fig.~\ref{fig:anc_admr_psitz}.
As $a$ increases, a peak develops near $\theta = 90^\circ$.
Although this peak remains weaker than that observed experimentally, its appearance indicates that a finite $z$-direction hopping of $\psi$ indeed produces the correct qualitative behavior.
Moreover, the peak at $\theta = 0^\circ$ is stronger than that at $\theta = 45^\circ$, consistent with experimental trends.
Achieving a more quantitative agreement with experiment will likely require a refined tuning of the Fermi pocket geometry, particularly near its corners.

\section{A brief introduction to the Yamaji effect.}\label{intro_yamaji}

\begin{figure}[t]
    \centering
    \includegraphics[width=0.96\linewidth]{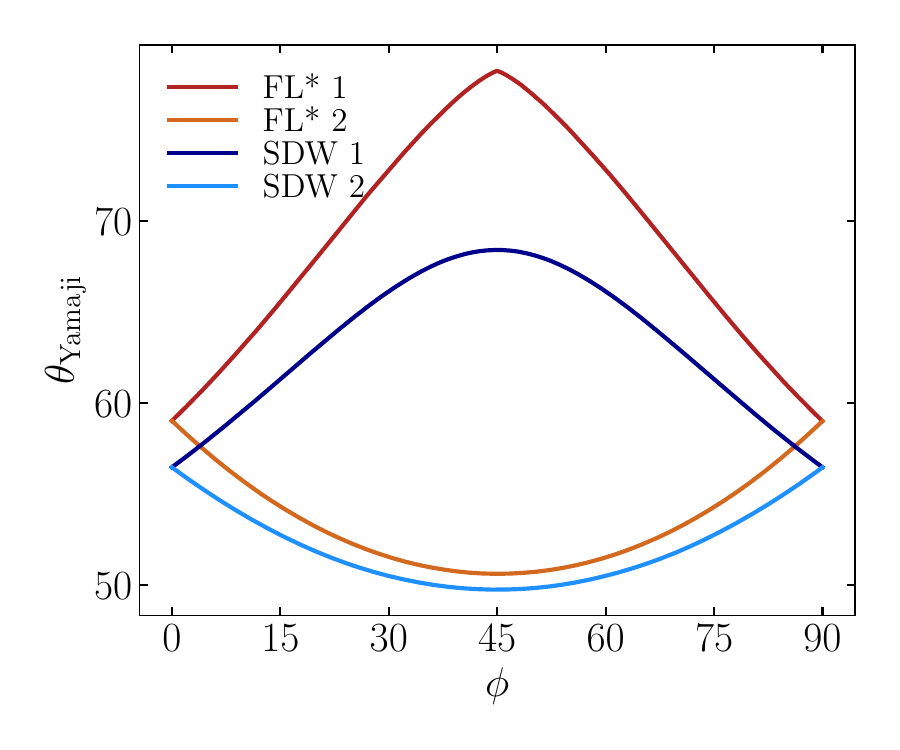}
    \caption{
        The Yamaji angles of a single pocket in the FL* and SDW states, directly estimated using the caliper momentum $k_{\mathrm{cal}}$ according to Eq.~\eqref{eqn:yamaji_angle}. 
        Here, FL*\ 1 and FL*\ 2 denote the two distinct pockets located in the first and second quadrants, respectively; SDW\ 1 and SDW\ 2 are defined analogously.
    }
    \label{fig:theta_yamajis}
\end{figure}

\subsection{Chamber's formula of conductivity}\label{chambers_formula}

We employ the semiclassical Boltzmann approach to calculate the conductivity. 
Within the relaxation-time approximation, the Boltzmann equation reads
\begin{equation}
    \frac{\partial f}{\partial t} + \mathbf{v} \cdot \nabla_{\mathbf{r}}f + \frac{\mathrm{d}\mathbf{k}}{\mathrm{d}t}\cdot \nabla_{\mathbf{k}} f = -\frac{\delta f}{\tau}\,,
\end{equation}
where $f(\mathbf{r},\mathbf{k},t)=f_0(\mathbf{k}) + \delta f(\mathbf{r},\mathbf{k},t)$ is the Fermi distribution of a semiclassical wave packet, and $f_0(\mathbf{k})$ is the equilibrium Fermi–Dirac distribution. 
The quasiparticle velocity is given by $\mathbf{v} = \nabla_{\mathbf{k}}\epsilon(\mathbf{k})$. 
In the presence of external electromagnetic fields, the momentum evolves as $\frac{\mathrm{d}{\bf k}}{\mathrm{d}t} = -e(\mathbf{E}+\mathbf{v}\times\mathbf{B})$. 
For a uniform external field, the dependence on $\mathbf{r}$ can be neglected, and the equation simplifies to 
\begin{equation}
    \frac{\partial f}{\partial t} + e (\mathbf{E} + \mathbf{v}\times \mathbf{B})\cdot \mathbf{v} \left(-\frac{\partial f}{\partial \epsilon}\right) = -\frac{\delta f}{\tau}\,.
\end{equation}
Its solution takes the form
\begin{equation}
    \delta f(\mathbf{k},t) = -e\mathbf{E}\cdot \mathbf{v}_{\mathrm{ret}}(\mathbf{k}(t),t)\tau(\mathbf{k}(t)) \left(-\frac{\partial f}{\partial \epsilon}\right),
\end{equation}
where the retarded velocity is defined as

\begin{equation}\label{eqn:retardvz}
     \mathbf{v}_{\mathrm{ret}} (\mathbf{k}(t),t) = \frac{1}{\tau(\mathbf{k}(t))}\int_{-\infty}^t d t' \mathbf{v}(\mathbf{k}(t')) e^{-\int_{t'}^t\frac{1}{\tau(\mathbf{k}(t''))}dt''}\,. 
\end{equation}
In the high-field regime where $\omega_c\tau \sim 1$, electrons undergo cyclotron motion along the Fermi surface, and the effect of $\mathbf{E}$ can be neglected in evaluating Eq.~\eqref{eqn:retardvz}. 

With the distribution function determined, the current density can be written as
\begin{widetext}
\begin{equation}
    \begin{aligned}
    J_i = -2e \int \frac{d^3k}{(2\pi)^3} f(\mathbf{k},0) v_i 
    = 2e^2 E_j \int \frac{d^3k}{(2\pi)^3} \left(-\frac{\partial f}{\partial \epsilon}\right)\int _{-\infty}^0 d t' v_i(\mathbf{k}(0)) v_j(\mathbf{k}(t')) e^{-\int_{t'}^{0}\frac{1}{\tau(\mathbf{k}(t''))}dt''}\,,
    \end{aligned}
\end{equation}
where the factor of 2 accounts for spin degeneracy. 
Here we have set $t=0$, since integration over $\mathbf{k}$ already averages over different times. 
This leads to the well-known Chamber’s formula for the conductivity:
\begin{equation}
     \sigma_{ij} = \frac{e^2 }{4\pi^3}\int_{\mathrm{FS}} d^2k \mathcal{D}(\mathbf{k})\int _{-\infty}^0 d t' v_i(\mathbf{k}(0)) v_j(\mathbf{k}(t')) 
     e^{-\int_{t'}^{0}\frac{1}{\tau(\mathbf{k}(t''))}dt''}\,,
\end{equation}
where $\mathcal{D}(\mathbf{k})$ is the local density of states, and $\mathbf{k}(t)$ parametrizes the position along the Fermi-surface orbit.
The subscript ``FS'' indicate integration over the full 3d Fermi surface. 
In all the calculations in this work, we assume $t_\perp$ is much smaller than the in-plane hopping, 
such that the effect of $v_z(B_x-B_y)$ can be neglected. 
All the cyclotron orbits share the same period in this condition.
We also assume $\mathcal{D}(\mathbf{k}) \approx \mathcal{D}_{2d}(\mathbf{k})$, with the later fully determined by the in-plane dispersion $\epsilon_{2d}(k_x, k_y)$. 
In the main text, we ignore the momentum dependence of $\tau(\mathbf{k})$ and approximate $\tau$ as the average scattering time.
\end{widetext}

\subsection{Yamaji effect}\label{yamaji-effect}

When $\omega_c \tau \sim 1$, electrons complete full periodic orbits along the Fermi surface. 
The velocity correlation term $v_i(\mathbf{k}(0))v_j(\mathbf{k}(t'))$ therefore averages over the entire orbit and cancels periodically under certain ``magic'' angles—this is the essence of the Yamaji effect.

As an example, consider a system with interlayer coupling $t_\perp$ and dispersion
\begin{equation}
    \epsilon_{3d}(\mathbf k) = \epsilon_{2d}(k_x,k_y) - 2t_\perp \cos(k_z c_{\mathrm{lat}})\,,
\end{equation}
where $\epsilon_{3d}(\mathbf{k})$ if the full 3d dispersion and $\epsilon_{2s}(k_x,k_y)$  is the in-plane dispersion, $c_{\mathrm{lat}}$ is the lattice constant along the $z$ direction.
When $t_\perp$ is much smaller than the in-plane hopping, all the cyclotron orbits share the same period, and $\mathcal{D}(\mathbf{k}) = \mathcal{D}_{2d}(\bfkt)$. 
The integration over the Fermi surface can replaced by an integration over $k_z$ and time $t$.
The conductivity along the $z$ direction is then
\begin{widetext}
\begin{equation}
\begin{aligned}
    \sigma_{zz} 
    =& \frac{e^3B\cos\theta}{4\hbar^2\pi^3}\int_{-\pi/c_{\mathrm{lat}}}^{\pi/c_{\mathrm{lat}}} dk_z \int_0^T dt \int_0^{\infty} d t' \, v_z(\mathbf{k}(t)) v_z(\mathbf{k}(t-t')) e^{-t'/\tau} \\
    =& \frac{t_\perp^2 c_{\mathrm{lat}}^2 e^3B\cos\theta}{\hbar^4\pi^3}\int_{-\pi/c_{\mathrm{lat}}}^{\pi/c_{\mathrm{lat}}} dk_z \int_0^T dt \int_0^{\infty} d t' 
    \sin(k_z c_{\mathrm{lat}} - k_\parallel(t) c_{\mathrm{lat}} \tan\theta) \sin(k_z c_{\mathrm{lat}} - k_\parallel(t-t') c_{\mathrm{lat}} \tan\theta) e^{-t'/\tau} \\
    =& \frac{t_\perp^2 c_{\mathrm{lat}} e^3B\cos\theta}{2\hbar^4\pi^3} \int_0^T dt \int_0^{\infty} d t' 
    \cos\!\left([k_\parallel(t) - k_\parallel(t-t')] c_{\mathrm{lat}} \tan\theta\right) e^{-t'/\tau}\,,
\end{aligned}
\end{equation}
\end{widetext}
where we assume a constant scattering time $\tau$, $T$ is the period of the quasi-particle cyclotron motion, and 
${k}_\parallel(t)$ denotes the in-plane component of the momentum along the cyclotron orbit projected to the in plane magnetic field direction $\mathbf{B}_\parallel$.

It is evident that the cosine term introduces an oscillatory dependence when $\tau$ is comparable to $T$. 
The oscillation period is controlled by the maximum momentum difference 
\(2k_{\mathrm{cal}} = \max\{|k_\parallel(t)-k_\parallel(t-t')|\}\), 
such that a larger $k_{\mathrm{cal}}$ corresponds to a smaller Yamaji peak angle.

Finally, in Fig.~\ref{fig:theta_yamajis} we show the Yamaji angles of FL* and SDW predicted directly from the caliper momentum $k_{\mathrm{cal}}$ according to Eq.~\eqref{eqn:yamaji_angle}. 
The results are consistent with the $c$-axis resistivity shown in the main text. 
It is clear that the SDW theory always predicts a smaller Yamaji angle than FL* along all the in-plane magnetic field direction $\phi$, with the difference most pronounced along $\phi=45^\circ$ for the first quadrant pockets. 

\bibliography{refs.bib}

\end{document}